\documentclass[aps,prd,showpacs]{revtex4}

\usepackage{amsmath,amsfonts,amssymb}
\usepackage[dvips]{graphicx}

\newtheorem{theorem}{Theorem}
\newtheorem{definition}{Definition}
\newcommand{\tr}{\mathop{\text{tr}}}
\newcommand{\ud}{\mathrm{d}}
\newcommand{\Real}{\mathop{\text{Re}}}

\begin{document}

\title{Dissipative or conservative cosmology with dark energy?} 
\author{Marek Szyd{\l}owski}
\email{uoszydlo@cyf-kr.edu.pl}
\affiliation{Astronomical Observatory, Jagiellonian University, Orla 171, 
30-244 Krak{\'o}w, Poland}
\affiliation{Marc Kac Complex Systems Research Center, Jagiellonian University, 
Reymonta 4, 30-059 Krak{\'o}w, Poland}
\author{Orest Hrycyna}
\email{hrycyna@kul.lublin.pl}
\affiliation{Department of Theoretical Physics, Faculty of Philosophy, 
The John Paul II Catholic University of Lublin, Al. Rac{\l}awickie 14, 
20-950 Lublin, Poland}
\date{\today}

\begin{abstract}
All evolutional paths for all admissible initial conditions of FRW cosmological
models with dissipative dust fluid (described by dark matter, baryonic matter
and dark energy) are analyzed using dynamical system approach. With that
approach, one is able to see how generic the class of solutions leading to the
desired property -- acceleration -- is. The theory of dynamical systems also
offers a possibility of investigating all possible solutions and their
stability with tools of Newtonian mechanics of a particle moving in a 
1-dimensional potential which is parameterized by the cosmological scale 
factor. We demonstrate that flat cosmology with bulk viscosity can be treated 
as a conservative system with a potential function of the Chaplygin gas type. 
We characterize the class of dark energy models that admit late time de Sitter
attractor solution in terms of the potential function of corresponding
conservative system. We argue that inclusion of dissipation effects makes the
model more realistic because of its structural stability. We also confront
viscous models with SNIa observations. The best fitted models are obtained by
minimizing the $\chi^{2}$ function which is illustrated by residuals and
$\chi^{2}$ levels in the space of model independent parameters. 
The general conclusion is that SNIa data supports the viscous model without the
cosmological constant. The obtained values of $\chi^{2}$ statistic are
comparable for both the viscous model and $\Lambda$CDM model. The Bayesian
information criteria are used to compare the models with different power law
parameterization of viscous effects. Our result of this analysis shows that
SNIa data supports viscous cosmology more than the $\Lambda$CDM model if the 
coefficient in viscosity parameterization is fixed. The Bayes factor is also 
used to obtain the posterior probability of the model. 

\end{abstract}

\pacs{98.80.Bp, 98.80.Cq}

\maketitle

\section{Introduction}
Several astronomical observations, since observations of distant supernova type
Ia \cite{riess:1998,perlmutter:1999} and cosmic microwave background anisotropy
measurements \cite{spergel:2003,tegmark:2004}, have indicated that our Universe is
currently undergoing an accelerating phase of expansion. If we assume that the
Friedmann equation with perfect fluid source is valid then our Universe is
in acceleration phase due to the presence of matter with negative pressure
violating the strong energy condition (dark energy). While there are several
candidates for description of dark energy like an evolving scalar field
(referred to as quintessence) \cite{ratra:1988}, generalized Chaplygin gas
\cite{gorini:2003,biesiada:2005} or phantom dark energy
\cite{caldwell:2002,dabrowski:2003} the most obvious candidate for such a
component of the dark energy seems to be the cosmological constant
\cite{weinberg:1989}. Hence we obtain the most simple explanation of current
universe which require two dark components, namely non-relativistic dust (dark
and baryonic matter) contributing about 30\% of the total energy density (which
is clustering gravitationally at small scales) and second component (dark
energy) which dominate at large scales and is smoothly distributed with
constant density. Unfortunately it is only some kind of effective theory which
offers description of present observations rater then its understanding. For
example it remains to understand why the value of cosmological constant
obtained from observations of SNIa is so small in comparison with the vacuum
energy (Planck mass scale). It is the main reason for looking for alternative
explanation of observational data of distant supernovae. There have appeared
many models that make use of new physics which may arise as a consequence of
embedding of our universe in a more dimensional universe which accelerates due to
some additional term in the Friedmann equation rather than dark energy
contribution \cite{deffayet:2002}.

In this work we discuss the possibility of the dark energy being represented
by viscous fluid which is characterized by the presence of bulk viscosity
coefficient  $\xi = -(1/3)\partial p_{\text{eff}}/\partial H$ in the effective
equation of state. Because we assume homogeneity and isotropy, only bulk
viscosity effects may be important under such symmetry condition. FRW models
filled with a non-ideal (dissipative) fluid were investigated in several papers
in which bulk viscosity was parameterized by an additional term in pressure $p \to
p_{\text{eff}} = p - 3 \xi H$, where $\xi$ is the bulk viscosity coefficient,
usually parameterized, following Belinskii, as power law $\xi(\rho)=\bar{\alpha}
\rho^{m}$, where $\rho$ is the energy density and $\bar{\alpha}$ and $m$ are
constants \cite{belinskii:1975,belinskii:1979,Chimento:1993,Elst:1994}. While
we know since the classical papers of Murphy \cite{Murphy:1973} and Heller
\cite{Heller:1973} that bulk viscosity effects can generate an accelerating
phase of expansion the idea that the present acceleration epoch is driven by some
kind viscous fluid is being explored recently \cite{Fabris:2005ts}. Wilson et al. 
\cite{Wilson:2006gf} discussed a cosmology in this cold dark matter particle 
decay into relativistic particles and argued such a decay could lead naturally 
to bulk viscosity modelled by effective negative pressure.

The main aim of the paper is to investigate the possibility of explaining the
present accelerating expansion of the Universe in terms of bulk viscosity with
the background of models with Robertson-Walker symmetry, basing on analogy with 
the cosmology with the Chaplygin gas and dynamical system methods. We show that the 
non-flat FRW dissipative models can treated as a perturbed dynamical systems of 
the Newtonian type. 

The Chaplygin gas conception is taken from aerodynamics. Kamenshchik at al. 
\cite{Kamenshchik:2000iv} noted that from this equation of state can 
interpolate two stages -- the stage of matter domination and the stage of dark 
energy domination. Then it was generalized to the form of the generalized 
Chaplygin gas $p = -A / \rho^{\alpha}$ by taking $\alpha$ different from one. 
The reason was to increase the degree of generalization without any physical 
justification. 

To have the physical meaning of this class of models we proposed to consider 
the FRW model with bulk viscosity which is formally indistinguishable from the 
flat FRW model with the generalized Chaplygin gas. The advantage of the FRW model 
with bulk viscosity is that its parameter possesses the physical interpretation 
-- dissipation in the sense of bulk viscosity although the integral of energy 
is conserved and this model belongs to the class of models called conservative 
models described in terms of the potential function.

One can distinguish between conservative and dissipative dynamical systems 
(divergence of the vector field is zero for conservative, larger then zero for 
dissipative). We show that the flat FRW with viscous fluid is conservative in 
the above sense. If we consider non flat models then the analogy between 
the flat FRW model with bulk viscosity and the Chaplygin gas model vanishes. 

In investigation of dynamical effect of bulk viscosity we use methods of
dynamical system theory which offers the possibility of investigating all
evolutional paths for all admissible initial conditions. In this approach the
key point is the visualization (geometrization) of dynamics with the help of
phase space. We consider dissipative cosmology (as well as conservative) in 
terms of a single potential function which determines the dynamics on a 
$2$-dimensional phase space $(a,\dot{a})$, where $a$ is the scale factor, 
$\dot{} \equiv {\ud}/{\ud t}$, and $t$ is the cosmological time.

Inspired by the fact that viscous fluid possesses negative pressure and in
flat models can be modelled as a Chaplygin gas we have undertaken the simple
task of studying the FRW viscous cosmology. The dissipative cosmological model have
at least a few significant advantages from the theoretical as well as
observational point of view:
\begin{enumerate}
\item{they can describe a smooth transition from a decelerated -- matter
dominated expansion of the Universe to the present epoch of cosmic
acceleration;}
\item{they offer the possibility of unification on phenomenological ground
both dark energy and dark matter;}
\item{they can be treated as a natural extension of the CDM models in which effects
of dissipation are unified; moreover from confrontation with SNIa data they are
a serious alternative to the concordance $\Lambda$CDM model;}
\item{in the flat cosmology they can be identified with conservative models
which are structurally stable, i.e. such that a small perturbation does not
change their global scenario.}
\end{enumerate}

Organization of the text is the following. In section \ref{sec:2} we describe
all classes of FRW conservative models with dark energy in terms of potential
function for fictitious particle moving on the constant energy level. In section
\ref{sec:3} the dynamics of conservative dark energy models with dissipation
is investigated in tools of qualitative theory of differential equation in the 
$2$-dimensional phase space. The full analysis of dynamics of viscous models in 
terms of perturbed conservative systems is performed for the constant bulk 
viscosity coefficient. We also define the distance the conservative and 
perturbative systems using the Sobolev metric. Section \ref{sec:4} contains 
some observational constraints from distant supernovae type Ia. These 
observational constraints indicate that unified model for dark energy and dark 
matter through the employment of dissipation is favored over the $\Lambda$CDM 
model by the Bayesian information criteria of model selection.

\section{Cosmological models with dark energy as a conservative system}
\label{sec:2}
The FRW dynamics of a broad class of dark energy models can be represented by
the motion of a fictitious particle in one dimensional potential $V(a)$, where $a$
is the scale factor. The equation of motion assumes the form of a $2$-dimensional 
dynamical system of a Newtonian type
\begin{equation}
\ddot{a} = -\frac{\partial V}{\partial a}(a) ,
\label{eq:1}
\end{equation}
where $V(a) = -\rho_{\textrm{eff}} a^{2}/6$ and effective energy density
satisfies the conservation condition
\begin{equation}
\dot{\rho}_{\textrm{eff}} = -3 H (\rho_{\textrm{eff}} + p_{\textrm{eff}}) ,
\label{eq:2}
\end{equation}
where $H=\dot{a}/a$ is the Hubble function and $a$ is the scale factor expressed in
units of its present value $a_{0}=1$. Therefore different dark energy universe
models filled with dust matter and dark energy satisfy the general form of the
equation of state
\begin{equation}
p_{X} = w_{X}(a) \rho_{X} ,
\label{eq:3}
\end{equation}
where 
\begin{equation}
\begin{array}{l}
p_{\text{eff}} = 0 + p_{X}, \\
\rho_{\text{eff}} = \rho_{\text{m},0} a^{-3} + \rho_{X,0} a^{-3} 
\exp{\Big(-3 \displaystyle{\int_{1}^{a} \frac{w_{X}(a)}{a} \ud a}\Big)},
\end{array}
\nonumber
\end{equation}
are the effective pressure and energy density, respectively. The index zero denotes
that the quantities are calculated at the present epoch. They can be classified in
terms of the potential function (see Table~\ref{tab:1}). In this table we used
redshift $z$ instead of the scale factor $a$ because of the relation 
$1 + z = a^{-1}$ and dimensionless density parameters for each component 
$\Omega_{i,0} = \rho_{i}/3H_{0}^{2}$.
\begin{table}
\caption{Different dark energy models in terms of the potential function
parameterized by the scale factor or redshift $z$ ($1+z=a^{-1}$).}
\begin{ruledtabular}
\begin{tabular}{|c|c|c|}
model & potential function & independent parameters \\
\hline \hline
$\Lambda$CDM model & $V(a(z))=-\frac{1}{2}\big\{ \Omega_{\text{m},0}(1+z) 
+ \Omega_{\Lambda,0}(1+z)^{-2} \big\}$ & $(H_{0},\Omega_{\text{m},0})$ $\sum_{i} \Omega_{i}=1$ \\
\hline
PhCDM model& & \\  (phantom $w_{X}=-4/3$) & $V(a(z))=-\frac{1}{2}\big\{ \Omega_{\text{m},0}(1+z) + \Omega_{ph,0}(1+z)^{-3} + \Omega_{k,0}\big\}$ & $(H_{0},\Omega_{\text{m},0},\Omega_{ph,0})$ \\
\hline
B$\Lambda$CDM model & & $n>3$; $\Omega_{\text{m},0},\Omega_{n,0}>0$; \\ bouncing with $\Lambda$ & $V(a(z))=-\frac{1}{2}\big\{ \Omega_{\text{m},0}(1+z) - \Omega_{n,0}(1+z)^{n-2} + \Omega_{\Lambda,0}(1+z)^{-2} + \Omega_{k,0}\big\}$ & $(H_{0},\Omega_{\text{m},0},\Omega_{n,0},\Omega_{\Lambda,0})$ \\
\hline
Randall-Sundrum & & \\
brane model with $\Lambda$ & $V(a(z))=-\frac{1}{2}\big\{ \Omega_{\text{m},0}(1+z) + \Omega_{d,0}(1+z)^{2} + \Omega_{\text{brane,0}}(1+z)^{4} + \Omega_{k,0}\big\}$ & $(H_{0},\Omega_{\text{m},0},\Omega_{d,0},\Omega_{\text{brane},0})$ \\
\hline
models with equation & & \\
of state $w_{X}=w_{0}+w_{1}z$ & $V(a(z))=-\frac{1}{2}\big\{ \Omega_{\text{m},0}(1+z) + \Omega_{X,0}(1+z)^{w_{0}+3(1-w_{1})} \exp{\big[3w_{1}z\big]}$ & $(H_{0},\Omega_{\text{m},0},\Omega_{X,0},w_{0},w_{1})$\\
linearized at $z=0$ & $+\Omega_{k,0}\big\}$& \\
\hline
Cardassian models & & \\
$3H^{2}=\rho + B \rho^{n}$ &$V(a(z))=-\frac{1}{2}\big\{ \Omega_{\text{m},0}(1+z) + \Omega_{\text{Card},0}(1+z)^{3n-2} + \Omega_{k,0}\big\}$ & $(H_{0},\Omega_{\text{m},0},\Omega_{\text{Card},0},n)$ \\
\hline
Sahni, Shtanov & $V(a(z))=-\frac{1}{2}\big\{ \Omega_{\text{m},0}(1+z) + \Omega_{k,0} + \Omega_{\sigma}(1+z)^{-2} + 2 \Omega_{l}(1+z)^{-2}$ & $\Omega_{\text{m},0} + \Omega_{k,0} + \Omega_{\sigma}+$\\
brane models & $\pm 2 \sqrt{\Omega_{l}}\sqrt{\Omega_{\text{m},0}(1+z)^{-1}+(1+z)^{-4}(\Omega_{\sigma}+\Omega_{l}+\Omega_{\Lambda,0})}\big\}$ & $2\sqrt{\Omega_{l}}\sqrt{1-\Omega_{k,0}+\Omega_{\Lambda,0}}=1$\\ 
 & & $(H_{0},\Omega_{\text{m},0},\Omega_{\sigma},\Omega_{l},\Omega_{\Lambda,0})$ \\
\hline
models with generalized & & \\
Chaplygin gas & $V(a(z))=-\frac{1}{2}\big\{ \Omega_{\text{b},0} 
+ \Omega_{\text{Ch},0}(1+z)^{-2}\bigg[ A_{s} 
+ (1-A_{s})(1+z)^{3(1+\alpha)}\bigg]^{\frac{1}{1+\alpha}}$ 
& $(H_{0},\Omega_{\text{Ch},0}, A_{s}, \alpha)$ \\
$p_{X}=-\frac{A}{\rho_{X}^{\alpha}}$ & $+\Omega_{k,0}\big\}$ & \\
and baryonic matter & & 
\end{tabular}
\end{ruledtabular}
\label{tab:1}
\end{table}

There are different advantages of representing the evolution of the Universe in
terms of a $2$-dimensional dynamical system of a Newtonian type. Firstly, we 
obtain unified description of the broader class of cosmological models with a 
general form of the equation of state. Secondly, the shape of potential which 
determines the critical points and their stability can be reconstructed 
immediately from the SN Ia data due to simple relation between the luminosity 
distance $d_{L}(z)$ and the Hubble function in the case of the flat model. 
Thirdly, the potential function can be useful as an instrument of probing of 
dark energy models from observations because as opposed to the coefficient of 
state $w_{X}(z)$ it is related to the luminosity distance $d_{L}(z)$ by a 
simple integral.

For our further analysis it is important that all these systems are
conservative ones. They are given on zero energy level which is conserved
\begin{equation}
\mathcal{H} = \frac{1}{2} p_{a}^{2} + V(a) \equiv 0,
\label{eq:4}
\end{equation}
where $p_{a}=\dot{a}$.

The dynamic of the models under considerations is reduced to the dynamical
system of the Newtonian form
\begin{equation}
\begin{array}{l}
\displaystyle{\dot{x} = y}, \\
\displaystyle{\dot{y} = - \frac{\partial V}{\partial x}(x)},
\end{array}
\label{eq:5}
\end{equation}
where $x=a/a_{0}$ and the above system has the first integral (energy) in the form
\begin{equation}
\displaystyle{\frac{y^{2}}{2}+V(x) \equiv 0}.
\label{eq:6}
\end{equation}
It is important that the critical points of (\ref{eq:5}) as well as their
character can be determined from the geometry of the diagram of the potential
function only.

The system (\ref{eq:5}) has critical points localized on $x$-axis because
$x=x_{0}$ and $(\partial V/\partial x)|_{x_{0}}=0$ at the critical point.
Therefore they are representing the static solution. The domain admissible for
classical motion is $\mathcal{D}_{0}=\big\{x \colon V(x) \le 0 \big\}$.

For the conservative system (\ref{eq:5}) it is useful to develop methods of
qualitative investigations of differential equation \cite{szydlowski:2006a}.
The main aim of this approach is to construct phase portraits of the system
which contain global information about the dynamics. The phase space $(x,y)$
offers a possibility of natural geometrization of the dynamical behavior. It 
is in simple $2$-dimensional case structurized by critical points or non-point 
closed trajectories (limit cycles) and trajectories joining them. Two phase 
portraits are equivalent modulo homeomorphism preserving orientation of the 
phase curves (or phase trajectories). From the physical point of view critical 
points (and limit cycles) are representing asymptotic states of the system or 
equilibria. Equivalently, one can look at the phase flow as a vector field
\begin{equation}
\boldsymbol{f}=\bigg[y,-\frac{\partial V}{\partial x}\bigg]^{T},
\label{eq:7}
\end{equation}
whose integral curves are the phase curves.

Thanks to Andronov and Pontryagin \cite{Andronov:1937}, the important idea of structural 
stability was introduced into the multiverse of all dynamical systems. A vector 
field, say $\boldsymbol{f}$, is a structurally stable vector field if there is 
an $\varepsilon > 0$ such that for all other vector fields $\boldsymbol{g}$, 
which are close to $\boldsymbol{f}$ (in some metric sense)
$\|\boldsymbol{f}-\boldsymbol{g}\|<\varepsilon$, $\boldsymbol{f}$ and
$\boldsymbol{g}$ are topologically equivalent. The notion of structural
stability is mathematical formalization of intuition that physically realistic
models in applications should posse some kind of stability, therefore small
changes of the r.h.s. of the system (i.e., vector field) does not disturb the
phase portrait. For example motion of pendulum is structurally unstable because
small changes of vector field (constructed from r.h.s. of the system) of
friction type $\boldsymbol{g}=[y,-\partial V/\partial x + k y]^{T}$
dramatically changes the structure of phase curves. While for the pendulum,
the phase curves are closed trajectories around the center, in the case of a 
pendulum with friction (constant) they are open spirals converging at the 
equilibrium after infinite time. We claim that the pendulum without friction 
is structurally unstable. Many dynamicists believe that realistic models of 
physical processes should be typical (generic) because we always try to convey 
the features of typical garden variety of the dynamical system. The exceptional 
(non-generic) cases are treated in principle as less important because they 
interrupt discussion and do not arise very often in applications 
\cite{Abraham:1992}.

In the $2$-dimensional case, the famous Peixoto theorem gives the 
characterization of the structurally stable vector field on a compact two 
dimensional manifold. They are generic and form open and dense subsets in the 
multiverse of all dynamical systems on the plane. If a vector field $f$ is not 
structurally stable it belongs to the bifurcation set.

The space of all conservative dark energy models can be equipped with the
structure of Banach space with the $C^{1}$ metric.

Let $V_{1}$ and $V_{2}$ be two dark energy models. Then $C^{1}$ distance
between them in the multiverse is
\begin{equation}
d(V_{1},V_{2}) = \max \Big\{\sup_{x \in E} | V_{1,x} - V_{2,x}|, 
\sup_{x \in E} |V_{1,xx}-V_{2,xx}|\Big\},
\label{eq:8}
\end{equation}
where $E$ is a closed subset of configuration space. Of course, multiverse of
all dynamical systems of a Newtonian type on the plane is infinite dimensional
functional space and the introduced metric is a so-called Sobolev metric.

While there is no counterpart of the Peixoto theorem in higher dimensions it can be
easy to test whether planar polynomial systems, like in considered case, have
structurally stable phase portraits. For this aim the analysis of behavior of
trajectories at infinity should be performed. One can simply do that using the
tools of Poincar{\`e} $S^{2}$ construction, namely by projection trajectories
from the center of the unit sphere $S^{2}=\big\{ (X,Y,Z) \in \mathbf{R}^{3} \colon
X^{2}+Y^{2}+Z^{2}=1 \big\}$ onto the $(x,y)$ plane tangent to $S^{2}$ at either
the north or south pole.

The vector field $f$ is structurally unstable if
\begin{enumerate}
\item{there are non-hyperbolic critical points on the phase portrait,}
\item{there is a trajectory connecting saddles on the equator of $S^{2}$.}
\end{enumerate}
In opposite cases if additionally the number of critical point and limit cycles
is finite, $f$ is structurally stable on $S^{2}$.

\begin{figure}
\begin{center}
\includegraphics[scale=0.85]{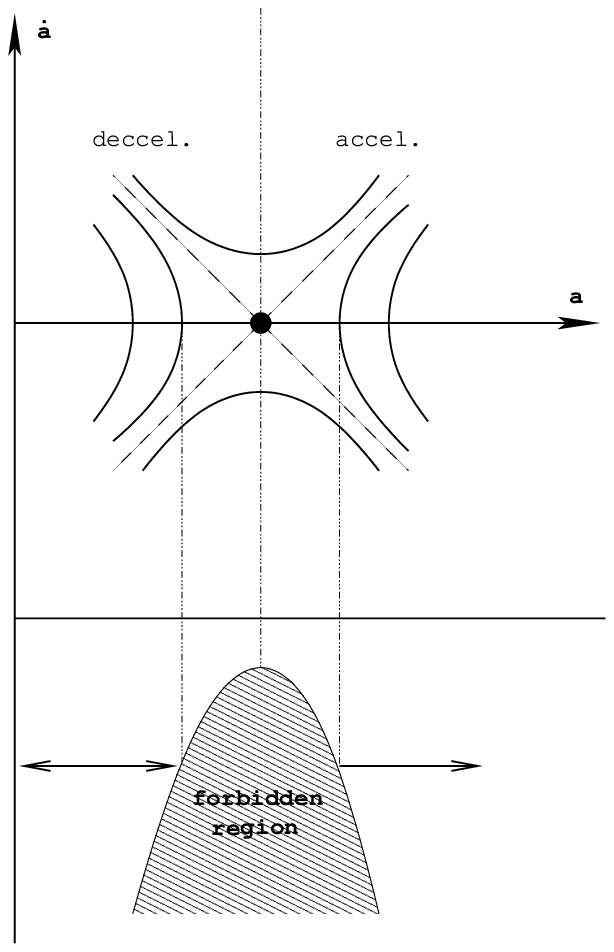}
\includegraphics[scale=0.85]{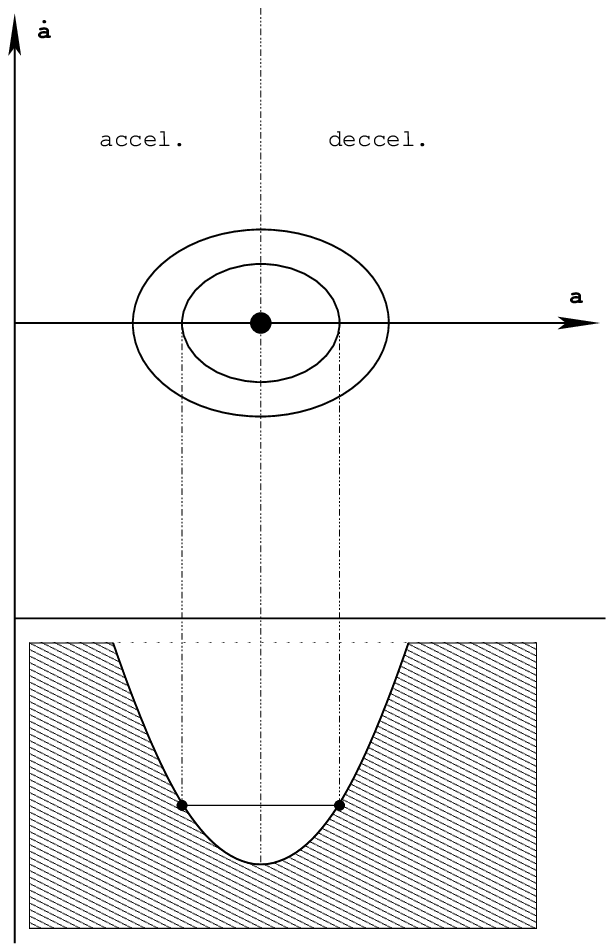}
\includegraphics[scale=0.85]{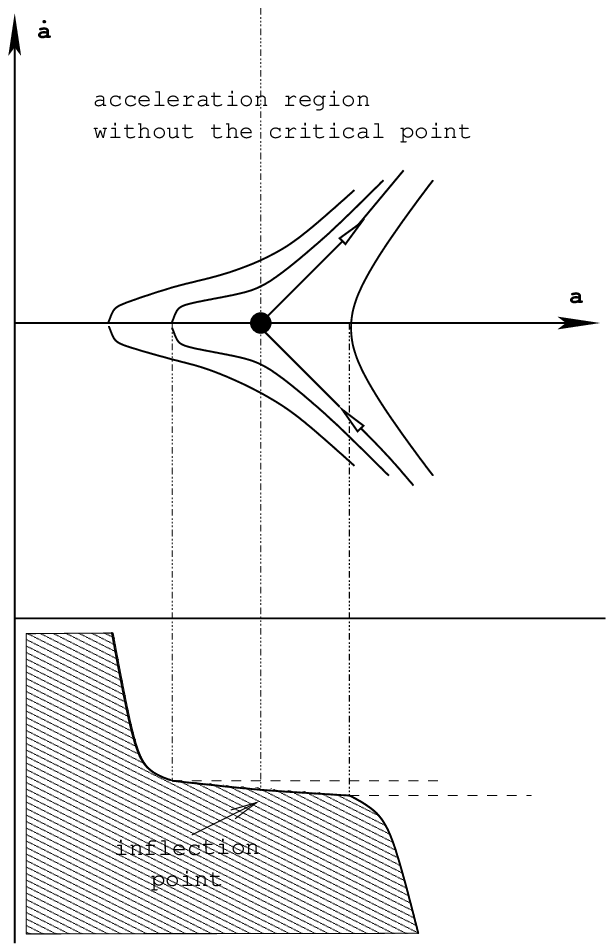}
\end{center}
\caption{Three types of behavior in the neighborhood of a critical point: a
saddle, a center and a cusp.}
\label{fig:1}
\end{figure}

Let us consider a $2$-dimensional dynamical system of a Newtonian type (\ref{eq:5}). 
There are three cases of behavior of the system admissible in the neighborhood of the
critical point $(x_{0},0) \colon -\partial V/ \partial x |_{x_{0}}=0$
\begin{itemize}
\item{If $(x_{0},0)$ is a strict local maximum of $V(x)$, it is a saddle point;}
\item{If $(x_{0},0)$ is a strict local minimum of $V(x)$, it is a center;}
\item{If $(x_{0},0)$ is a horizontal inflection point of $V(x)$, it is a cusp.}
\end{itemize}
All these cases are illustrated in Fig.~\ref{fig:1}.

It is a simple consequence of the fact that characteristic equation for
the linearization matrix at the critical point $(\tr{A} = 0 )$ is $\lambda^{2} +
\det{A} = 0$, where $\det{A} = \partial^{2} V/\partial x^{2} |_{x_{0}}$.
Therefore the eigenvalues are real of opposite sign for saddle point, and for
centers -- if they are non-hyperbolic critical points -- purely imaginary and 
conjugated.

Finally, if the set of all conservative dark energy models has potential
function $V(x)$ such that the number of static critical points and cycles is
finite and there are no trajectories connecting saddles then $f \in
C^{r}(\mathcal{M})$, $r \ge 1$ contains open, dense subset of all smooth vector
fields of class $C^{r}$ on the phase space.

\begin{figure}
\begin{center}
\includegraphics[scale=1]{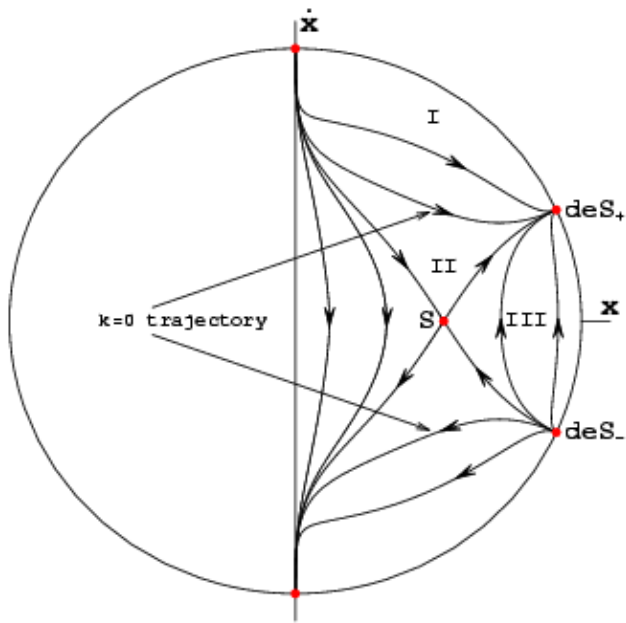}
\end{center}
\caption{The phase portrait for the $\Lambda$CDM model. On the phase portrait
we have a single saddle point in the finite domain and four critical points located
on the circle at infinity. They represent initial singularity
$(x=0,\dot{x}=\infty)$ or de Sitter Universe deS. \\ The trajectory of the flat
model ($k=0$) divides all models in to two disjoint classes: closed and open.
The trajectories situated in the region II confined by the upper branch of the
$k=0$ trajectory and by the separatrix going to the stable de Sitter node
$\textrm{deS}_{+}$ and by the separatrix going from the initial singularity to
the saddle point $S$ correspond to the closed expanding universe. The
trajectories located in the regions I and II (which corresponds to the open
universes) are called inflectional. Quite similarly, the trajectories situated in
the region III correspond to the closed universes contracting from the unstable
de Sitter node towards the stable de Sitter node. The trajectories running in
this region describe the closed bouncing universes.}
\label{fig:2}
\end{figure}

In Fig.~\ref{fig:2} the phase portrait for the $\Lambda$CDM model is presented
on a compactified projective plane by circle at infinity. Of course it is
structurally stable. Therefore, following Peixoto theorem, it is generic in
the multiverse $\mathcal{M}$ of all dark energy models with $2D$ phase space
because such systems form open and dense subsets.

\begin{figure}
\begin{center}
\includegraphics[scale=0.5,angle=270]{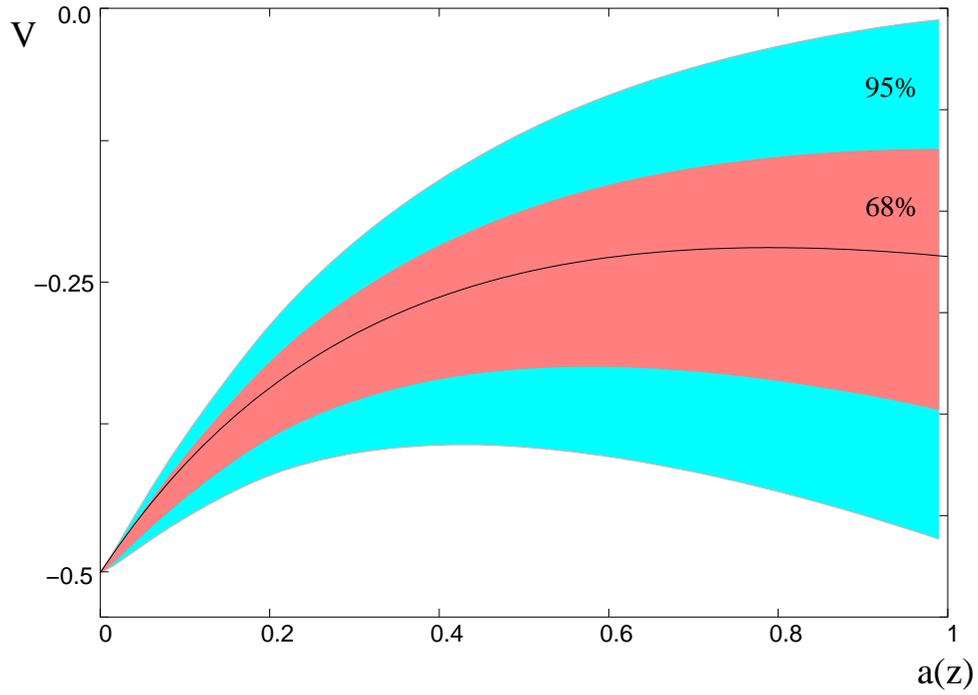}
\end{center}
\caption{The potential function as a function of the scale factor expressed in
its present value $a_{0}=1$ for the reconstructed best fit model is given by
the solid line. The confidence regions $1\sigma$ and $2\sigma$ are drawn around
it. The phase portrait obtained from this potential (best fit) is equivalent to
the $\Lambda$CDM model (see Fig.~\ref{fig:2}). The value of the redshift 
transition estimated from SNIa data (the Gold Riess sample) is about $0.38$ 
(see \cite{Czaja:2004}).}
\label{fig:3}
\end{figure}

It is also interesting that the phase space of dark energy model can be
reconstructed from SN Ia data set (Gold Riess sample) and is topologically
equivalent to the $\Lambda$CDM model. Fig.~\ref{fig:3} represents the potential
function
\begin{equation}
V(a(z)) = -\frac{1}{2}(1+z)^{2} \Bigg[\frac{\ud}{\ud z} \frac{d_{L}(z)}{1+z}\Bigg]^{-2},
\label{eq:9}
\end{equation}
reconstructed from $d_{L}(z)$ relation -- luminosity distance $d_{L}$ as a
function of redshift $z: 1+z=a^{-1}$. Such a reconstruction is possible due to
the existence of the universal formula
\begin{equation}
\frac{d_{L}(z)}{1+z} = \int \frac{\ud z'}{H(z')},
\label{eq:10}
\end{equation}
for the flat model.

In principle it is good news for the $\Lambda$CDM model and others which are
close to it in the multiverse in the metric sense.

\begin{figure}
\begin{center}
\includegraphics[scale=1]{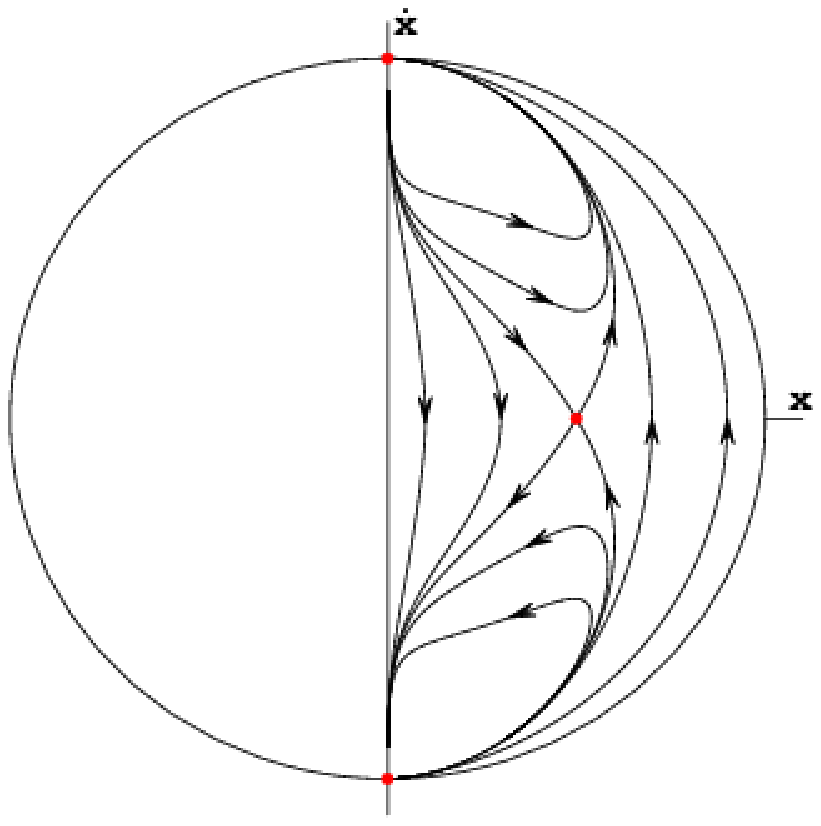}
\includegraphics[scale=1]{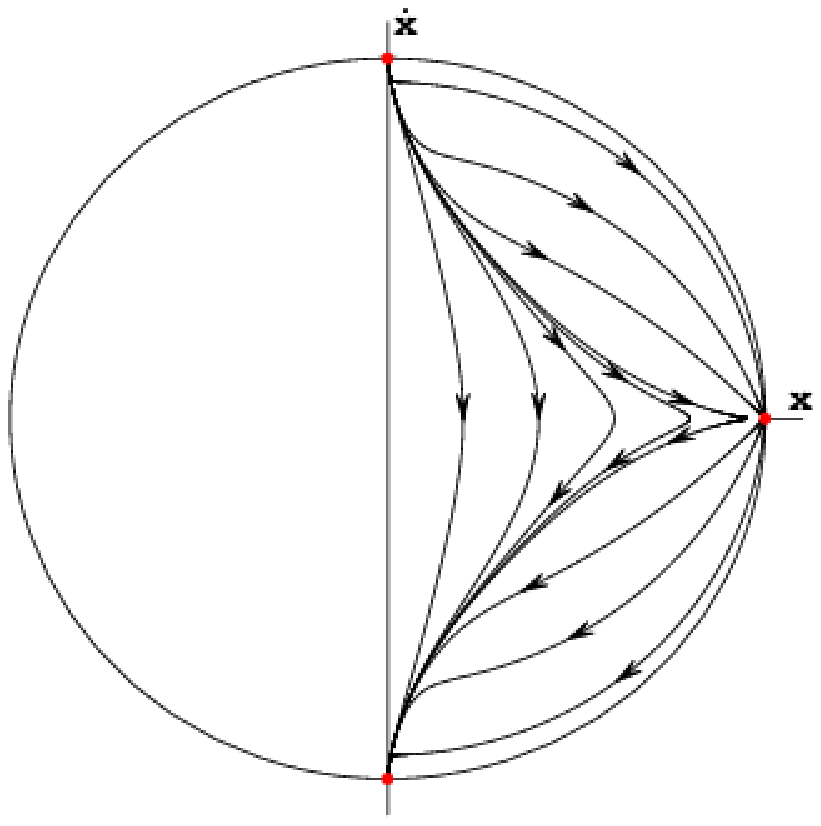}
\end{center}
\caption{The phase portraits for the Cardassian model $n=-1$ (left) and $n=2$
(right) (see Table \ref{tab:1}). The Cardassian model with $n=-1/3$ corresponds
to the phantom model with $w_{X}=\textrm{const}<-1$ (generally if we consider
dust model equation of state $p_{\textrm{card}}=(n-1)\rho_{\textrm{card}}$
reproduces the Cardassian term).}
\label{fig:4}
\end{figure}

On the phase portrait the phase space of the Cardassian model is presented 
(Fig.~\ref{fig:4}). In the case of single fluid, the presence of
additional terms in the modified Friedmann equation is equivalent to extra
contribution to the effective energy density in the standard Friedmann
equation. On the phase portrait there is a degenerated point
$x=\infty,y=0$ located at the infinity.
Let us note that there is no de Sitter attractor at late times. This obstacle 
was addressed in Ref.~\cite{hao:2003,lazkoz:2005}.

\begin{figure}
\begin{center}
\includegraphics[scale=1]{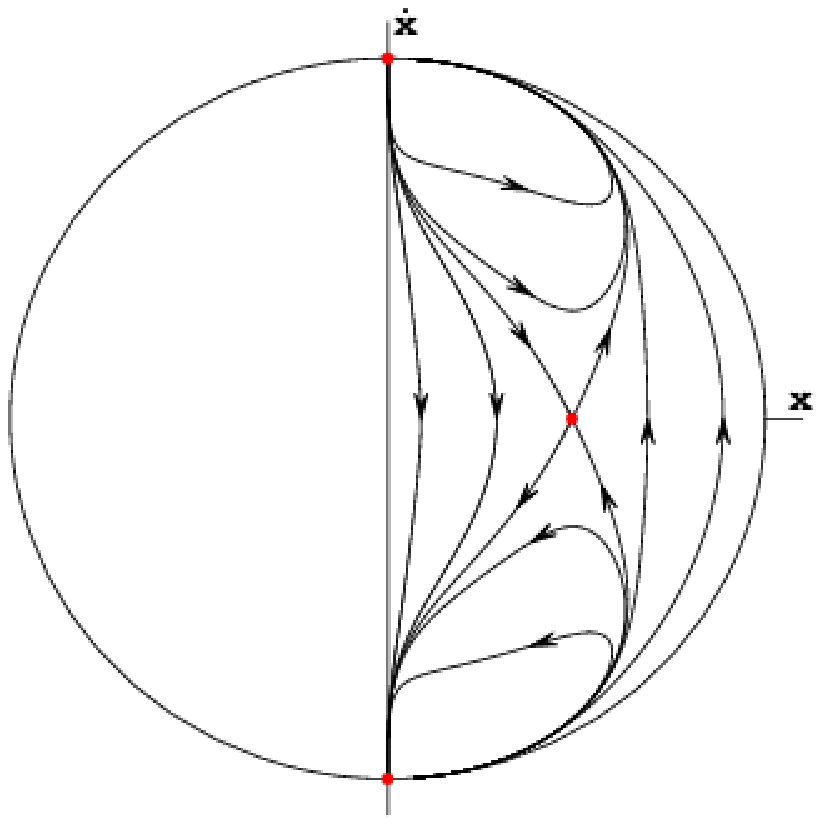}
\end{center}
\caption{The phase portrait for the phantom model. While at finite domain we
find only a saddle point like for the $\Lambda$CDM model, at infinity, the 
trajectory is reaching the degenerate point $x=0$, $\dot{x}=\infty$. Note that 
this degeneration can be removed be redefinition of the phase variables and
reparameterization of time. Then we obtain other system which is topologically
equivalent to the $\Lambda$CDM model (see Fig.~\ref{fig:2}).}
\label{fig:5}
\end{figure}

Analogous situation appears if we consider negative value of the exponent $n$ 
in the Cardassian models. It is well known that if we consider dust matter only
Cardassian models with $n < 0$ are equivalent to phantom models. The phase
portrait for the phantom model ($p_{X}=-(4/3) \rho_{X}$) with dust is
illustrated in Fig.~\ref{fig:5}.
In both cases, Fig.~\ref{fig:4} and Fig.~\ref{fig:5}, the systems are
structurally unstable due to the presence of degenerate critical points at
infinity. The models with de Sitter phase of evolution at late time are
distinguished in the multiverse of dark energy models.

Our main objection addressed to the Cardassian models is that they do not offer
description of dark energy epoch in terms of a structurally stable model.
Cardassian cosmologies belong to the non-generic, in our terminology, class of
fragile model of the present Universe. Similar models form the bifurcation
set in the multiverse of dark energy models.

\begin{figure}
\begin{center}
\includegraphics[scale=1]{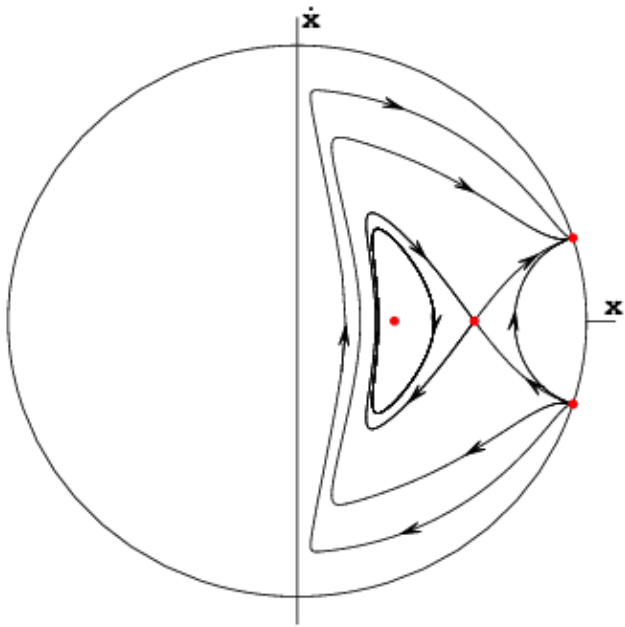}
\end{center}
\caption{The phase portrait for bouncing (in the contemporary sense see Table
\ref{tab:1}) cosmological model with $\Lambda$. On the $x$-axis, there are two 
critical points: center and saddle point. The global phase portrait (in
contrast to the $\Lambda$CDM model) is structurally unstable because of the presence 
of a non-hyperbolic critical point -- center. Note that all models (closed, open 
and flat) undergo a bounce, i.e., they begin their contraction from the unstable 
de Sitter node, reach a point of minimal contraction and begin expansion toward 
the stable node $\text{deS}_{+}$.}
\label{fig:6}
\end{figure}

There is also another example of the fragile model used in the cosmological
applications -- the bouncing model. Such a model recently appeared in the context
of semi-classical description of the quantum evolution of the Universe in the
framework of loop quantum gravity. In this scenario singularity separates
classically allowed regions \cite{Ashtekar:2005}. The phase portrait of
bouncing cosmological model is shown in Fig.~\ref{fig:6}. In such a cosmology
one encounters a bouncing phase of evolution instead of big bang during which the
potential function has a minimum. It is structurally unstable due to the presence
of non-hyperbolic critical point on the phase portrait. In the next section we
find some perturbed system which modifies this and makes it structurally
stable.

\section{Class of dissipative dark energy models}
\label{sec:3}
Applying the analogy with pendulum with friction, we would like to incorporate
dissipative processes into cosmological model with dark energy. As a matter of
fact, dissipative processes should be present in any realistic theory of the
Universe \cite{prisco:2000}.

The simplest way to include bulk viscosity effects is through Eckart's theory
\cite{eckart:1940}, which postulates that the bulk viscous pressure is
proportional to the expansion. Many authors used that theory to investigate
bulk viscosity on the evolution of the Universe
\cite{Murphy:1973,belinskii:1975}.

Basing on Israel and Stewart theory \cite{israel:1976,israel:1979} of
truncated viscosity, Belinskii \cite{belinskii:1979} studied effects of
viscosity on cosmological models in terms of effective pressure
\begin{equation}
p_{\textrm{eff}} = p - 3 \xi (\rho) H,
\label{eq:11}
\end{equation}
where $p$ is the pressure of a fluid filling the Universe, and $3
\xi(\rho)=\bar{\alpha} \rho^{m}$ is the viscosity coefficient parameterized by
energy density $\rho=\rho_{\text{eff}}$.

In the context of viscous cosmology many authors used dynamical system methods
in investigations of the stationary states and their stability
\cite{coley:1994}. Several authors have studied cosmological models with
viscosity \cite{maartens:1996}, which are also interesting in the context of
inflation \cite{oliveira:1998}. It is also interesting that viscosity may
describe phenomenologically quantum effects \cite{barrow:1988}.

In this section the effects of dissipation are treated as a small perturbation
of conservative dark energy models. We then investigate the influence of
dissipation on global dynamics of fragile dark energy models.

Let us start with the definition.
\begin{definition}
A multiverse of the FRW dissipative models with dark energy is defined as a space 
of all $2$-dimensional perturbed dynamical systems of a Newtonian type
\begin{equation}
\left\{ \begin{array}{l}
\displaystyle{\dot{x} = y} , \\
\displaystyle{\dot{y} = -\frac{\partial V}{\partial x}(x) + \bar{\alpha} \rho^{m}(x,y) y}.
\end{array} \right.
\label{eq:12}
\end{equation}
\end{definition}
We have used Belinskii parameterization of the viscosity
$3\xi(\rho)=\bar{\alpha} \rho^{m}$ and the first integral $\rho - 3 H^{2}=3
k/x^{2}$ or $\rho_{\text{eff}}=3 H^{2}$ is preserved also in the case of
viscous cosmology. If $\bar{\alpha}=0$ we obtain the standard conservative FRW
cosmology.

It is useful to distinguish two special cases.

\subsection{Flat models (k=0)}
In this case the dynamical effects of viscosity are equivalent to the effects
of Chaplygin gas \cite{Fabris:2005ts,Szydlowski:2006ay}. If we put $p_{\text{eff}}=\gamma
\rho - 3 \alpha \rho^{m} H$ then from the conservation condition $\dot{\rho}=-3
H (\rho + p_{\textrm{eff}})$ we obtain the relation
\begin{equation}
\rho(x) = \Bigg(\frac{A}{1+\gamma} + \frac{B}{x^{3(1+\gamma)(1+\alpha)}}\Bigg)^{\frac{1}{1+\alpha}},
\label{eq:13}
\end{equation}
where $A$ and $B$ are integration constants, and
$\rho(x=1)=\rho_{0}=\frac{A}{1+\gamma}+B$.

In the special case of $\alpha=0$ $(m=-1/2)$ and $\gamma=0$ (dust) the
corresponding equation relating energy density to the scale factor for
Chaplygin gas can be recovered. Therefore all the flat models with viscosity
following Belinskii parameterization can by represented as conservative
systems. To construct the phase space portrait it is useful to represent
the potential function $V(x)$ in the similar way as for generalized Chaplygin
gas for which the equation of state assumes the following form
\begin{equation}
p = - \frac{A}{\rho^{\alpha}}, \nonumber
\end{equation}
and energy density can be parameterized by $\rho_{0}$ and $A_{s}$
\begin{equation}
\rho(x) = \rho_{0} \Bigg( A_{s} + \frac{1-A_{s}}{x^{3(1+\gamma)(1+\alpha)}} \Bigg)^{\frac{1}{1+\alpha}},
\label{eq:14}
\end{equation}
where $c^{2}_{s}=\alpha A_{s}$ is the squared velocity of sound. Therefore the
potential function for the system is given by
\begin{equation}
V(x) = - \frac{\rho x^{2}}{6} = - \frac{1}{2} \Omega_{\text{visc},0} x^{2} 
\Bigg( A_{s} + \frac{1-A_{s}}{x^{3(1+\gamma)(1+\alpha)}} \Bigg)^{\frac{1}{1+\alpha}},
\label{eq:15}
\end{equation}
where $1+\alpha=1/2 - m$, $A_{s}=\frac{A}{1+\gamma}$, 
$\Omega_{\text{visc},0}=\frac{\rho_{0}}{3 H_{0}^{2}}$.

\begin{figure}
\begin{center}
a)\includegraphics[scale=1]{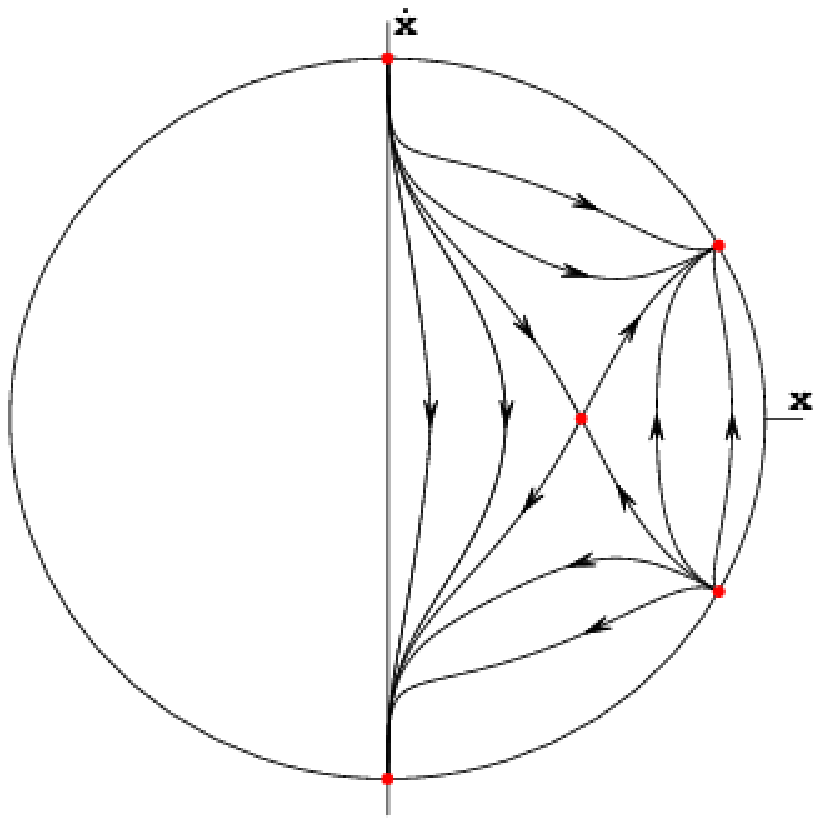}
b)\includegraphics[scale=1]{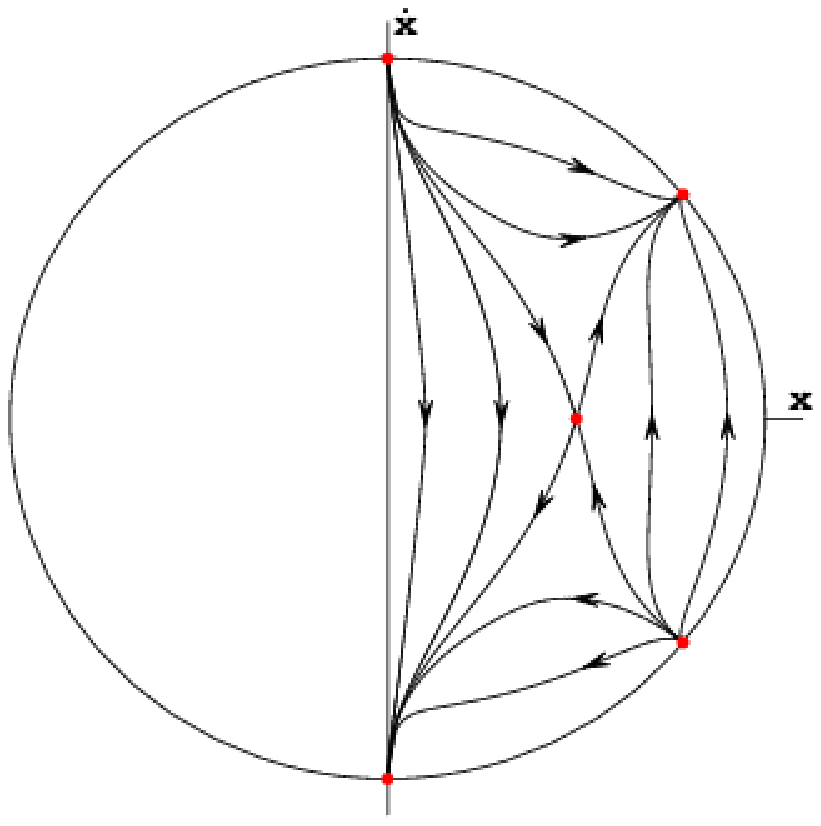}

c)\includegraphics[scale=1]{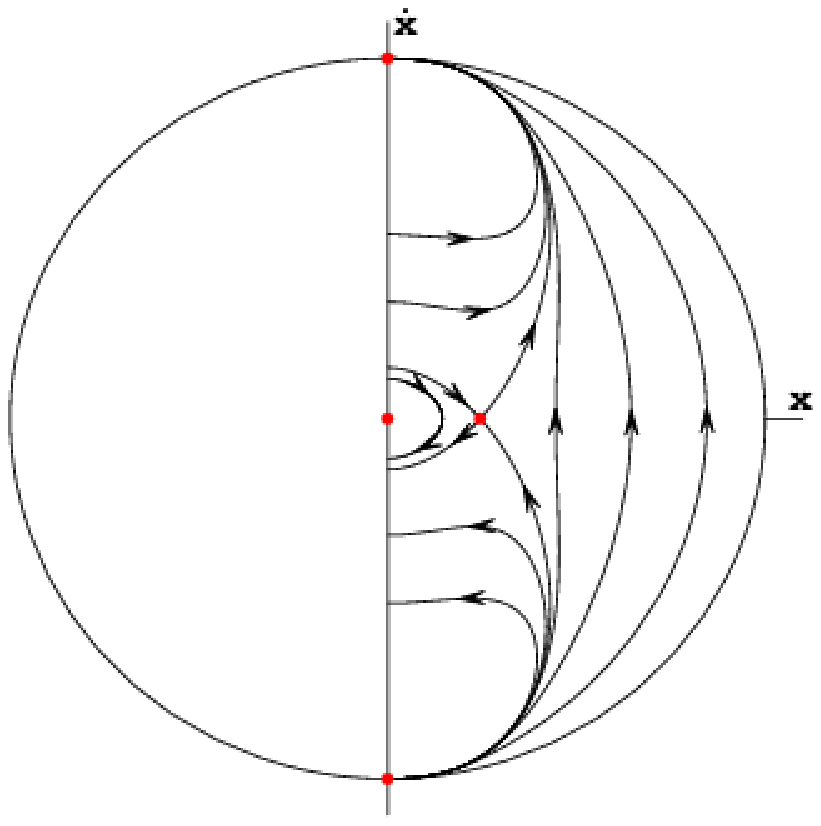}
\end{center}
\caption{The phase portraits for the viscous cosmological models. a)
$\gamma=0$, $\alpha=0$ ($m=-1/2$), $A_{s}=0.3$, b) $\gamma=1/3$, $\alpha=1/2$
($m=-1$), $A_{s}=0.5$, c) $\gamma=-4/3$, $\alpha=0$ ($m=-1/2$), $A_{s}=-0.95$.
Note that both models presented in Fig. a) and b) are topologically
equivalent.}
\label{fig:7}
\end{figure}

\begin{figure}
\begin{center}
d)\includegraphics[scale=1]{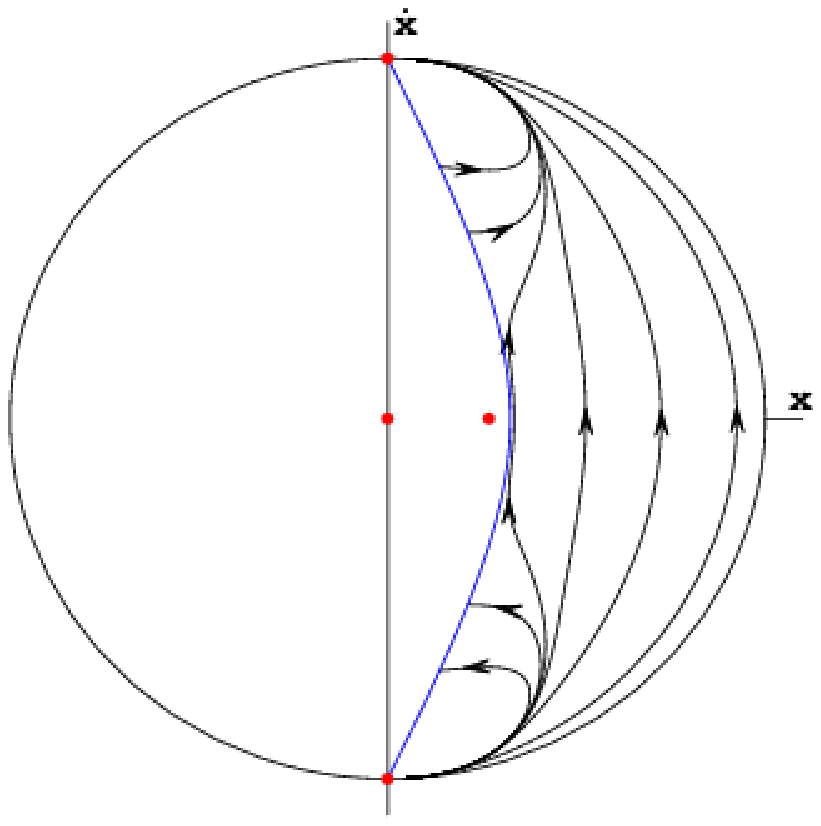}
e)\includegraphics[scale=1]{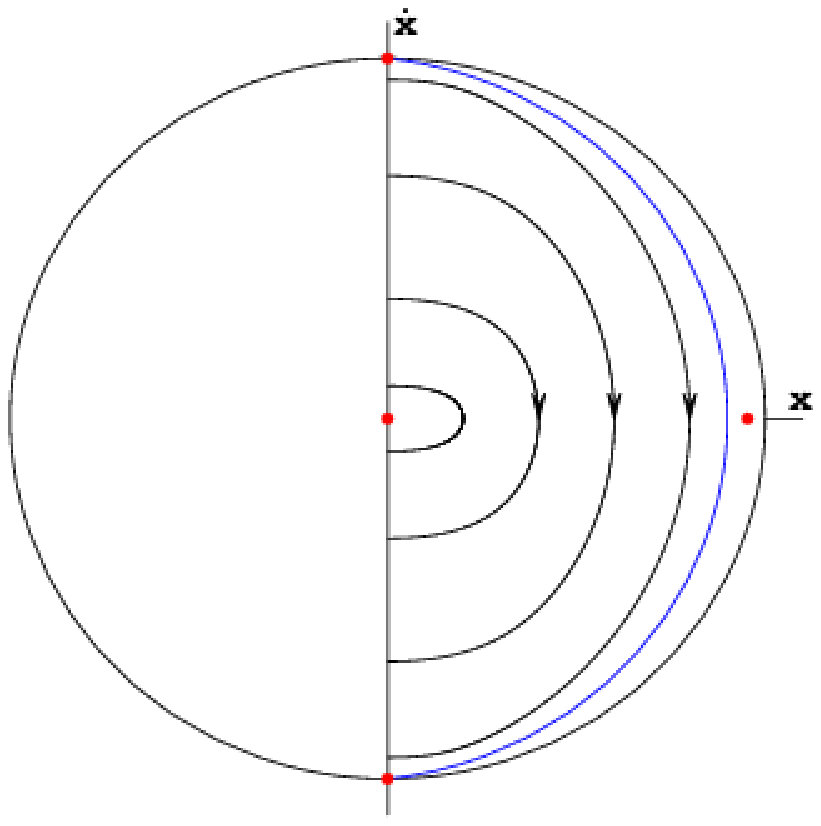}
\end{center}
\caption{The phase portraits for the viscous cosmological models. d)
$\gamma=-4/3$, $\alpha=1/2$ ($m=-1$), $A_{s}=-0.5$, and the solid (blue in el.
version) line denotes the limit $x=x_{\textrm{min}}$, e) $\gamma=-4/3$,
$\alpha=-3/2$ ($m=1$), $A_{s}=-0.5$, the solid line denotes the limit
$x=x_{\textrm{max}}$.}
\label{fig:7a}
\end{figure}

In Figs.~\ref{fig:7} and ~\ref{fig:7a} the phase
portraits for different parameter $m$ and for different equation of state
coefficient $\gamma$ are illustrated. Therefore the effect of viscosity in the
perfect fluid is equivalent to the effect of Chaplygin gas in the case of flat model.

Note that for phantom cosmology ($\gamma<-1$) $A_{s}$ can be either negative or
positive but if $A_{s}$ is negative, the expression in the parentheses should be
positive. Therefore we obtain following inequalities which restrict
$x=x_{\textrm{min}}$ or $x=x_{\textrm{max}}$
$$
x = \left\{
\begin{array}{lcl}
< \big(\frac{A_{s}-1}{A_{s}}\big)^{\frac{1}{3(1+\gamma)(1+\alpha)}} & \textrm{if} & (1+\gamma)(1+\alpha)>0,\\
> \big(\frac{A_{s}-1}{A_{s}}\big)^{\frac{1}{3(1+\gamma)(1+\alpha)}} & \textrm{if} & (1+\gamma)(1+\alpha)<0.\\
\end{array} \right.
$$
Therefore for $A_{s}<0$ ($1+\gamma<0$) $x<x_{\textrm{max}}$ if only $\alpha<-1$
($m>1/2$) and $x>x_{\textrm{min}}$ if only $\alpha>-1$ ($m<1/2$).

\begin{figure}
\begin{center}
\includegraphics{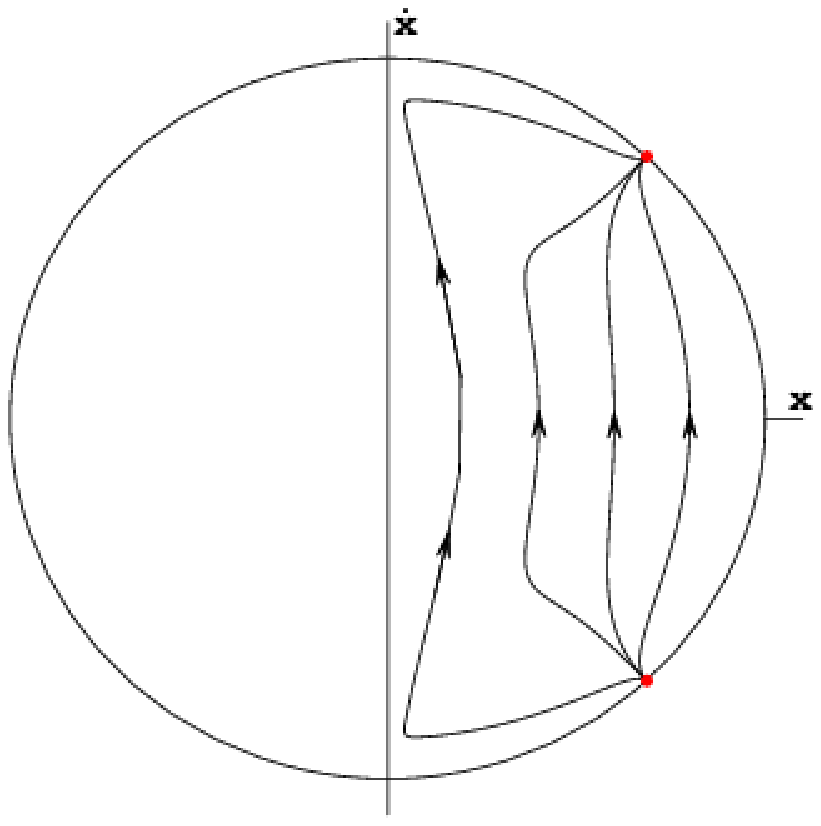}
\includegraphics{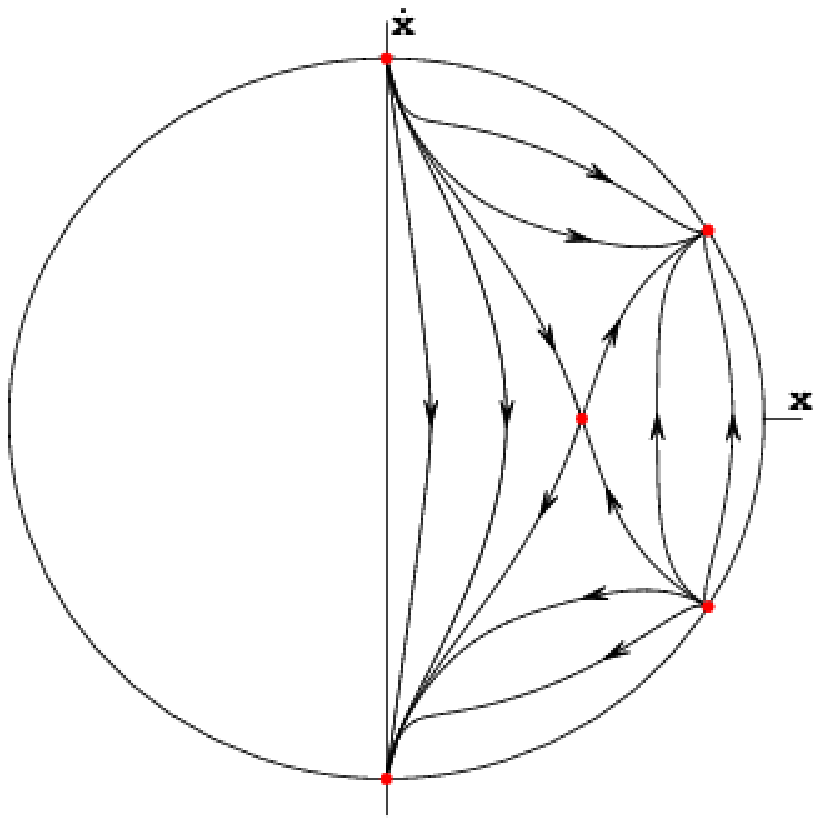}
\end{center}
\caption{The bouncing models with viscosity for potential (\ref{eq:16}) with
$A_{s}=0.75$. $\Omega_{\text{visc},0}>1$ (left) and $\Omega_{\text{visc},0}<1$ 
(right). Note that viscosity makes the global phase portraits of bouncing 
models structurally stable.}
\label{fig:8}
\end{figure}

If we assume that $\gamma > -1$, then the Chaplygin gas interpolates between the state of
positive cosmological constant and state of matter domination $\rho \propto
a^{-3(1+\gamma)}$ at the early epoch. Hence the viscosity effects can shift
structurally unstable model into open dense set. Fig.~\ref{fig:8} shows
the influence of viscosity on bouncing models for which the potential function
contains additional contribution which can unify matter $(p=0)$ and dark
energy.

If we postulate the existence of negative contribution to the effective energy
density coming, for example, from quantum effects like in Vandersloot's
approach \cite{Vandersloot:2005,Singh:2005}, then the potential function assumes the
following form
\begin{equation}
V(x) = -\frac{1}{2}\Bigg\{ \Omega_{\text{visc},0} x^{2} \bigg(A_{s} 
+ \frac{1-A_{s}}{x^{3(\frac{1}{2}-m)}}\bigg)^{\frac{1}{\frac{1}{2}-m}} 
- \frac{1}{2} \Omega_{lq,0} x^{-4} \Bigg\}.
\label{eq:16}
\end{equation}
where $\Omega_{\text{visc},0} - \Omega_{lq,0}=1$ and then the cosmological model
exhibits characteristic bounce phase \cite{Hossain:2005eu,szydlowski:2005}.

It is characteristic for loop quantum cosmology that initial singularity is
replaced with bounce because of quantum effects which are manifested by
presence of negative energy contribution to the $H^{2}(a)$ relation. In this
picture space-time does not end at singularities and quantum regime can offer
a bridge between vast space-time regions which are classically unrelated (see
\cite{Ashtekar:2005}).

Unfortunately, in the phase space we obtain a non-hyperbolic critical point 
around the bounce which makes the system structurally unstable. In 
Fig.~\ref{fig:8} it is illustrated how small viscosity effects (or equivalently 
the Chaplygin gas) can dramatically change global phase portraits and makes the 
corresponding system (with potential (\ref{eq:16})) structurally stable.

\subsection{Case of constant viscosity (m=0)}

Let us start now the analysis of the effects of viscosity in the case of 
constant viscosity. In this special case the equation of motion assumes a very 
simple form
\begin{equation}
\left\{ \begin{array}{l}
\displaystyle{\dot{x} = y} , \\
\displaystyle{\dot{y} = -\frac{\partial V}{\partial x}(x) + \alpha y}.
\end{array} \right.
\label{eq:17}
\end{equation}
One can calculate the distance from a dissipative system (\ref{eq:17}) to a
conservative one. Then the perturbation vector is $$\delta = [0, \alpha y
]^{T},$$ and distance $$ d(cons,dissip) = \max_{(x,y) \in E}\big\{|\alpha
y|,\alpha\big\}=\alpha \max_{(x,y) \in E}\big\{|y|,1\big\},$$ where $E$ is a
compact region of the phase space in which we compare models, which we chose as
the region $|y| < 1$. Hence the  parameter $\alpha$ will measure the distance
from a viscous perturbed system to a conservative system, in $C^{1}$ metric.
It is interesting that the exact solution of the system (\ref{eq:17}) with constant
viscosity can be simply obtained, without any assumption about the curvature,
if we know the corresponding solution without the viscosity. The following
theorem establishes this fact.
\begin{theorem}
Let $t=\phi(a)$ be a solution of the conservative system with dark energy
$$\dot{a}=(-2V(a))^{-1/2}.$$ It can be determined from the relation
$$t(a)=\int^{a}\frac{\ud x}{\sqrt{-2V(x)}}=\phi(a).$$ Then solution of the
type
\begin{equation}
\exp{[\alpha t]} = \phi(a),
\label{eq:18}
\end{equation}
will be the solution with constant viscosity $\alpha$.
\end{theorem}

To prove this fact, it is sufficient to check that solution of the above type
will satisfy equation $$\ddot{a}=-\frac{\partial V}{\partial a} + \alpha
\dot{a},$$ which is equivalent to equation (\ref{eq:17}).

One can consider now the properties of the phase plane of the dissipative dark
energy models. We are especially interested whether bouncing models can be
shifted to the generic set of the multiverse.

\begin{figure}
a)\includegraphics[scale=1]{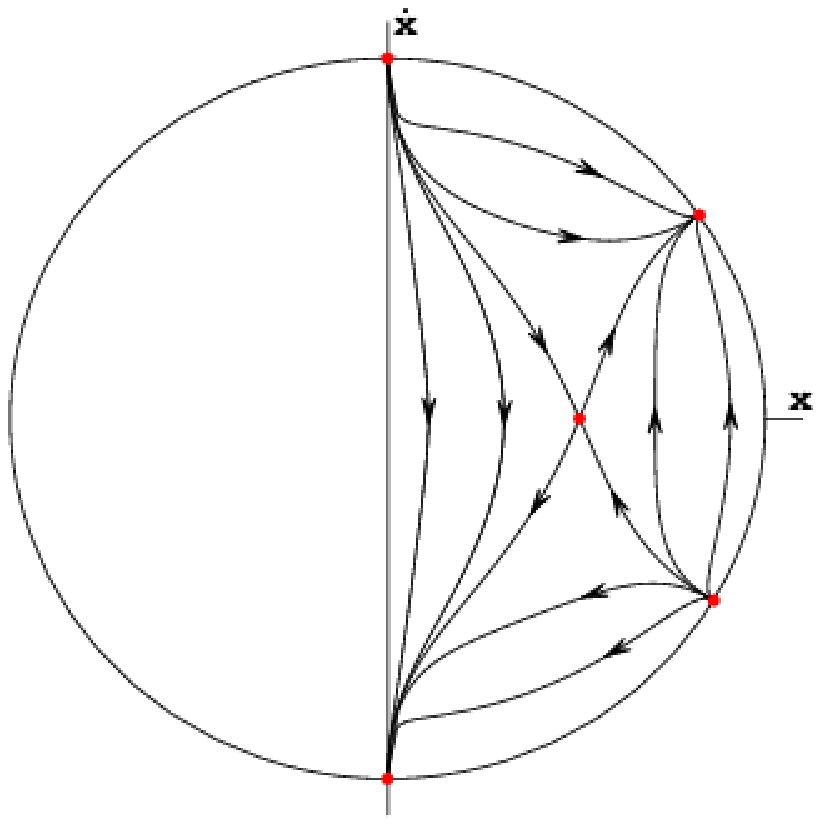}
b)\includegraphics[scale=1]{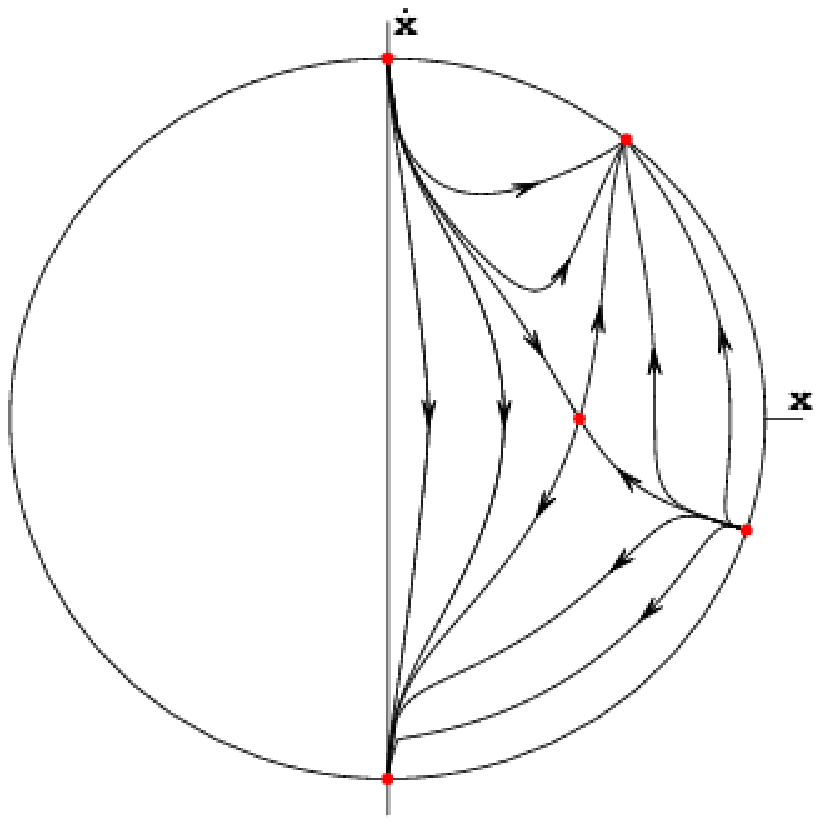}

c)\includegraphics[scale=1]{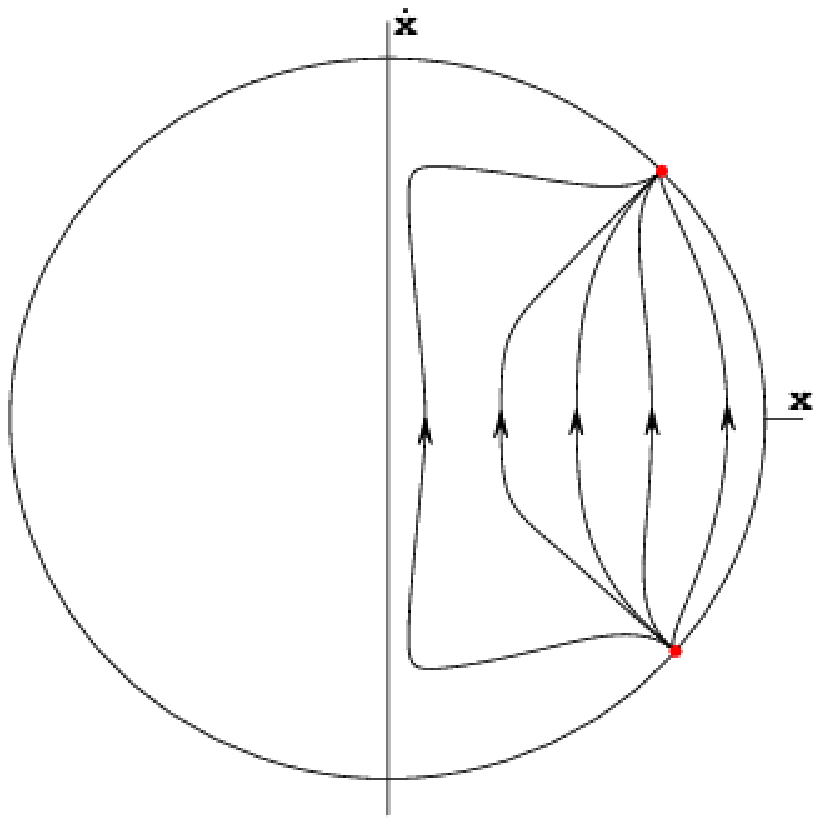}
d)\includegraphics[scale=1]{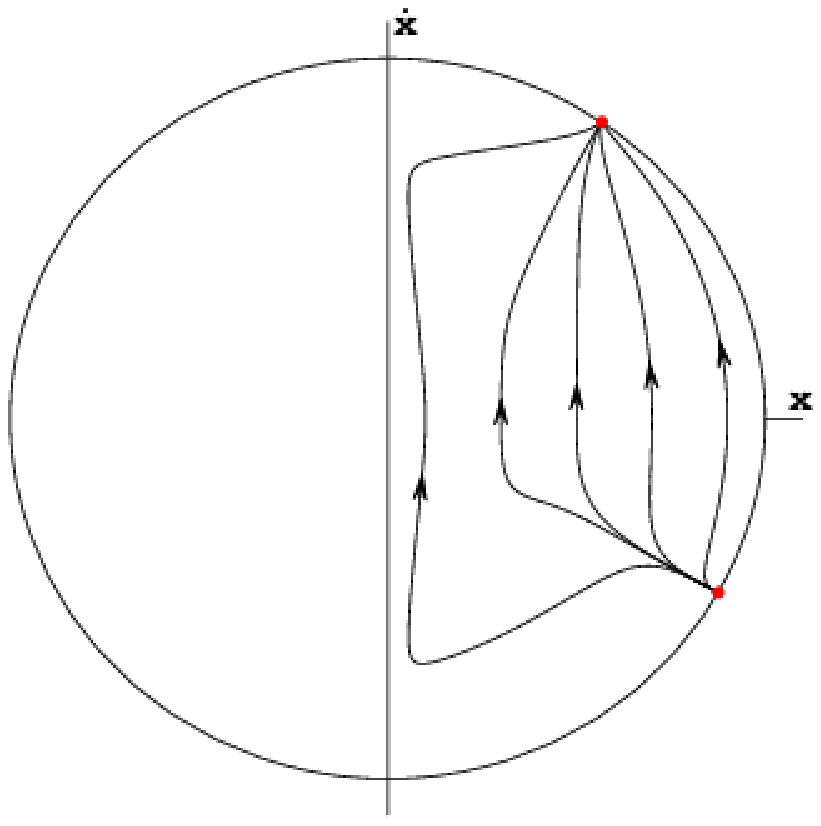}
\caption{The phase portrait of the system (\ref{eq:22}) for the bouncing
$\Lambda$CDM model with a constant viscosity coefficient which assumes different
values. Cases: a) $\Omega_{\Lambda,0}=0.4$, $\Omega_{\text{m},0}=0.3$, $\alpha=0.1$,
b) the same but with $\alpha=0.9$; c) $\Omega_{\Lambda,0}=0.8$,
$\Omega_{\text{m},0}=0.3$, $\alpha=0.1$, d) the same as previous but with
$\alpha=0.9$. Note that all phase portraits are structurally stable due to
viscosity perturbation and that symmetry $\dot{x} \to -\dot{x}$ can be broken
if viscosity effects are included. It is because entropy is changing due to the
presence of dissipative processes.}
\label{fig:9}
\end{figure}

The location of all critical points at finite domain coincides with that of the case
without dissipation effects because all should be situated on the $x$--axis
(static critical point). Their character is determined by the linearization
matrix
\begin{equation}
A = \left[ \begin{array}{cc}
0 & 1 \\
\displaystyle{-\frac{\partial^{2} V}{\partial x^{2}}} & \alpha
\end{array} \right]_{(x_{0},0)}.
\label{eq:19}
\end{equation}
Because $\tr{A}=\alpha > 0$ all the critical points are unstable. The
characteristic equation assumes the following form
\begin{equation}
\lambda^{2} - \alpha \lambda + V_{xx}(x_{0})=0.
\label{eq:20}
\end{equation}
Therefore the eigenvalues of the linearization matrix (\ref{eq:19}) are real of
the same signs if $V_{xx}(x_{0})>0$ and opposite signs if $\lambda_{1}
\lambda_{2} = V_{xx}(x_{0})<0$. They are real if $\Delta = \alpha^{2} -
4V_{xx}(x_{0})$ is positive, or imaginary if $\Delta$ is negative. In any case
$\Real{\lambda} =\alpha$ is different from zero, i.e., all critical points are
hyperbolic. Note that if $V_{xx}(x_{0}) < 0$ then always $\Delta > 0$, i.e.,
eigenvalues are real of opposite signs (they are representing saddles). If
$V_{xx}(x_{0}) > 0$ (like in the neighborhood of bounce) then eigenvalues are real
and positive ($\lambda_{1} + \lambda_{2}= \alpha$ and $\lambda_{1} \lambda_{2}
> 0$) or imaginary (if $V_{xx}(x_{0}) > \alpha^{2}/4$). This corresponds to the
presence of an unstable node or an unstable focus on phase portraits, respectively.
Because both unstable nodes are topologically equivalent to the unstable
focus, we obtain generic phase portraits for the dissipative bouncing model which
are demonstrated in Fig.~\ref{fig:9}.

Let us now present some general properties of the dissipative FRW models. The 
equation of motion is in the form
\begin{equation}
\left\{ \begin{array}{l}
\displaystyle{\dot{x} = y \equiv P(x,y)} , \\
\displaystyle{\dot{y} = -\frac{\partial V}{\partial x}(x) + \alpha y \equiv Q(x,y)}.
\end{array} \right.
\label{eq:17a}
\end{equation}
The critical points at finite domain (only repellors) are determined from
$$
y_{0} = 0, \qquad \bigg(\frac{\partial V}{\partial x}\bigg)_{x_{0}}=0.
$$
System (\ref{eq:17a}) linearized around the static critical point $(x_{0},0)$ has the form
$$
\left\{ \begin{array}{l}
\displaystyle{(x-x_{0})\dot{} = \bigg(\frac{\partial P}{\partial x}\bigg)_{(x_{0},0)}(x-x_{0}) + \bigg(\frac{\partial P}{\partial y}\bigg)_{(x_{0},0)}(y-y_{0})} , \\
\displaystyle{(y-y_{0})\dot{} = \bigg(\frac{\partial Q}{\partial x}\bigg)_{(x_{0},0)}(x-x_{0}) + \bigg(\frac{\partial Q}{\partial y}\bigg)_{(x_{0},0)}(y-y_{0})},
\end{array} \right.
$$
or
$$
\left\{ \begin{array}{l}
\displaystyle{\dot{x} = y}, \\
\displaystyle{\dot{y} = -\bigg(\frac{\partial^{2} V}{\partial x^{2}}\bigg)_{(x_{0},0)}(x-x_{0})}.
\end{array} \right.
$$
It is useful to shift the critical point to the origin of a new coordinate system labelled as $(X,Y)$, say $x \to X = x-x_{0}$, $Y=y$. Then
$$
\left\{ \begin{array}{l}
\displaystyle{\dot{X} = Y } , \\
\displaystyle{\dot{Y} = -\bigg(\frac{\partial^{2} V}{\partial X^{2}}\bigg)_{(0,0)}X},
\end{array} \right.
$$
or $\dot{\boldsymbol{X}}=A \boldsymbol{X}$.
Let us consider the special case of a conservative system for $\alpha=0$. The solutions of the system are
\[
X=C_{1} \exp{[\lambda_{1}t]} + C_{2} \exp{[\lambda_{2}t]},
\]
where $C_{1}$ and $C_{2}$ are constants and $\lambda_{1}$, $\lambda_{2}$ are eigenvalues of the linearization matrix.

There are two types of the solutions:
\begin{enumerate}
\item{solution of a saddle point if eigenvalues are real of opposite signs $\lambda_{1,2}= \pm \sqrt{-\det{A}}=\pm \sqrt{-\big(\frac{\partial^{2} V}{\partial X^{2}}\big)_{(0,0)}}$,}
\item{solution of a center type when eigenvalues are imaginary $\lambda_{1,2}= \pm i \sqrt{-\det{A}}=\pm i \sqrt{\big(\frac{\partial^{2} V}{\partial X^{2}}\big)_{(0,0)}}$.}
\end{enumerate}
Finally we obtain corresponding solution
$$
x-x_{0} = C_{1} \exp{[\sqrt{\bigg(-\frac{\partial^{2} V}{\partial x^{2}}\bigg)_{x=x_{0}}} t]} + C_{2} \exp{[-\sqrt{\bigg(-\frac{\partial^{2} V}{\partial x^{2}}\bigg)_{x=x_{0}}} t]},
$$
or
$$
x-x_{0} = C_{1} \cos{\sqrt{\bigg(\frac{\partial^{2} V}{\partial x^{2}}\bigg)_{x=x_{0}}} t} + C_{2} \sin{\sqrt{\bigg(\frac{\partial^{2} V}{\partial x^{2}}\bigg)_{x=x_{0}}} t}.
$$

In the case of dissipative ($\alpha =$ const) dynamical system the characteristic equation assumes the form (\ref{eq:20}) and solutions are $\lambda_{1,2} = - \frac{\alpha \pm \sqrt{\Delta}}{2}$ and $\lambda_{1} \lambda_{2} = \det{A} = \big(\frac{\partial^{2} V}{\partial x^{2}}\big)_{x_{0}}$. In this case the type of the critical points (if they exist) will depend on both the sign of discriminant $\Delta = \alpha^{2} -
4V_{xx}(x_{0})$ and sign of $V_{xx}(x_{0})$ at the critical points, namely
\begin{itemize}
\item{if $V_{xx}(x_{0})<0$, i.e. $V(x)$ has a maximum then critical points are 
always saddles,}
\item{if $V_{xx}(x_{0})>0$, i.e. $V(x)$ has a minimum at $x_{0}$ then character 
of the critical points will depend on the value of $\alpha$ such that:}
\begin{itemize}
\item{if $\alpha > 2 \sqrt{V_{xx}(x_{0})}$ then we have unstable nodes,}
\item{if $0 < \alpha <  2 \sqrt{V_{xx}(x_{0})}$ then we obtain unstable focus.}
\end{itemize}
\end{itemize}
Therefore in $\alpha$ is sufficiently small then unstable focus is present on 
the phase portrait instead of center in the conservative case. If $\alpha$ is 
sufficiently large then we obtain unstable node at the minimum of the diagram 
of the potential function. Note that except of degeneration case of center all 
admissible critical points are structurally stable. The focus is topologically 
equivalent to node.

Note that the presence of dissipative effects enlarge the accelerating domain
occupied by accelerating cosmological models because of relation
\begin{equation}
-\frac{\partial V}{\partial x}(x) + \alpha y > 0,
\label{eq:21}
\end{equation}
and now this region depends additionally on the coordinate $y$. Note that at
the critical point relation (\ref{eq:21}) is always valid (like in
\cite{Vandersloot:2005}).

If we substitute in (\ref{eq:17}) the form of the potential function for
bouncing cosmology with $\gamma=0$ (dust) and $n=4$, then the corresponding
dynamical system has the form
\begin{equation}
\left\{ \begin{array}{l}
\displaystyle{\dot{x} = y} , \\
\displaystyle{\dot{y} = -\frac{1}{2} \Omega_{\text{m},0} x^{-2} 
+ \Omega_{n,0} x^{-3} + \Omega_{\Lambda,0} x + \alpha y}.
\end{array} \right.
\label{eq:22}
\end{equation}
Rescaling the time variable we can make r.h.s. of the system (\ref{eq:22})
polynomial. The simple time transformation given by $\tau \mapsto \eta \colon
\ud \tau/x^{3} = \ud \eta$ give rise to its regularization. Such a choice of
new time variable does not change the phase portrait because our system has
been autonomous since the very beginning. This operation is equivalent to
multiplication of the r.h.s. by $x^{3}$ so that the system can be considered
in the $\eta$ time, because $x>0$ and $\eta$ is a monotonous function of any
time $\tau$. Finally we obtain
\begin{equation}
\left\{ \begin{array}{l}
\displaystyle{\frac{\ud x}{\ud \eta} = y x^{3}} , \\
\displaystyle{\frac{\ud y}{\ud \eta} =  -\frac{1}{2} \Omega_{\text{m},0} x 
+ \Omega_{n,0}  + \Omega_{\Lambda,0} x^{4} + \alpha y x^{3}},
\end{array} \right.
\label{eq:23}
\end{equation}
where $\Omega_{\text{m},0}-\Omega_{n,0}+\Omega_{\Lambda,0}=1$.

Note that reparameterization of cosmological time is equivalent to the choice of the
corresponding lapse function $N(t)$ in such a way that $\tau=\int N(t)\ud t$.
If we consider particle like representation of the dynamics then we obtain
classical mechanics with lapse in the H.-J. Schmidt terminology
\cite{Schmidt:1996}.

The degeneration of the dynamics at infinity for phantoms can be removed in a
simple way by introducing a new positional variable $\bar{x}$ such that $$ x
\mapsto \bar{x} = x^{\frac{3}{2}}$$ (in general $\bar{x}=x^{-(1+3w)/2}$) and
reparameterizing the time variable
$$
\begin{array}{rl} 
\tau \mapsto \eta \colon & \frac{3}{2} \bar{x}^{\frac{1}{3}} \ud t = \ud \eta, \\
\textrm{or}: & \frac{3}{2} x^{\frac{1}{2}} \ud t = \ud \eta.
\end{array}
$$
Of course the new time variable is a monotonous function of the original time
$\tau$. Hence we obtain the new system in which the original variable $y$ 
preserves its original sense
$$
\left\{ \begin{array}{l}
\displaystyle{\frac{\ud \bar{x}}{\ud \eta} = y} , \\
\displaystyle{\frac{\ud y}{\ud \eta} =  -\frac{1}{3} 
\left( -\Omega_{\text{m},0} \bar{x}^{-\frac{5}{3}} + 3 \Omega_{\text{ph}} \bar{x} \right) 
= - \frac{\partial V}{\partial \bar{x}}}. 
\end{array} \right.
$$
New potential function assumes the form 
$$ V(\bar{x}) = -\frac{1}{2}\bigg\{ \Omega_{\text{m},0} \bar{x}^{-\frac{2}{3}} 
+ \Omega_{\text{ph}} \bar{x}^{2} + \Omega_{k,0}\bigg\}.$$
The above system is of Newtonian type and has a first integral in the energy
conservation form $$ \frac{y^{2}}{2}+V(\bar{x})=0,$$ where now $y=\frac{\ud
\bar{x}}{\ud \eta}$.

It is easy to check that the kinetic energy form $\frac{1}{2}(\frac{\ud
\bar{x}}{\ud \eta})^{2} = \frac{1}{2}(\frac{\ud x}{\ud t})^{2}$ is preserved
under nonlinear rescaling transformation $x \mapsto \bar{x}$ and time
reparameterization. The form of the potential function $V(\bar{x})$ is
\textit{de facto} the same as the original $V(x)$ in which instead of $x$ we
operate on $\bar{x}$ as the positional variable.

From the first integral we obtain that for large $\bar{x}$ we have
$$\frac{y}{\bar{x}}=\sqrt{\Omega_{ph}}$$ because material contribution is 
negligible. It means that asymptotically the trajectories reach the point which lies on
the circle at infinity and on the straight line $y \propto \bar{x}$. The
corresponding dynamical system on the plane $(\bar{x},y)$ is of course
equivalent to the $\Lambda$CDM model phase portrait although we must remember that
the attractor at infinity does not represent the de Sitter universe. Therefore 
the obtained global phase portrait for phantom cosmology is structurally stable 
in a similar way as in the case of the $\Lambda$CDM model.

The analogous ``regularization procedure'' which we perform for phantom with
$w=-\frac{4}{3}$ can be realized in a more general case for any $w$. It is
sufficient to choose
$$
\begin{array}{rl}
x \mapsto \bar{x} \colon & \bar{x}=x^{-\frac{1+3w}{2}},\\
\tau \mapsto \eta \colon & \ud \eta = -(\frac{1+3w}{2}) x^{-\frac{3}{2}(1+w)} \ud t.\\
\end{array}
$$
The corresponding potential function is $$ V(\bar{x}) = -\frac{1}{2}\bigg\{
\Omega_{\text{m},0} \bar{x}^{-\frac{2}{3}} + \Omega_{w,0} \bar{x}^{2} + \Omega_{k,0}
\bigg\}. $$ Therefore phantom cosmology in which a weak energy condition is
violated forms an open and dense subsets in the multiverse of all dark energy
models. Also big-rip singularities which are attributed to these classes of models are
a generic features of their long term behavior.

The universe is accelerating in the region of phase space determined by the
condition
\begin{equation}
-\Omega_{\text{m},0} x + 2 \bigg\{\Omega_{n,0}  + \Omega_{\Lambda,0} x^{4} + \alpha y x^{3}\bigg\}>0,
\label{eq:24}
\end{equation}
and finally region of accelerating expansion is given by 
$$y>\frac{1}{\alpha x^{3}} \bigg\{\frac{1}{2}\Omega_{\text{m},0} x 
- \Omega_{n,0}  - \Omega_{\Lambda,0} x^{4}\bigg\} $$

Therefore all terms apart from the matter term $\Omega_{\text{m},0}$ act toward 
the acceleration of the Universe.

\section{Viscous cosmology tested by distant supernovae type Ia}
\label{sec:4}

Using the energy conservation condition we can obtain a simple relation
$$\rho(a)=\bigg[\frac{A}{1+\gamma}+\frac{B}{a^{3(1+\gamma)(\frac{1}{2}-m)}}
\bigg]^{\frac{1}{\frac{1}{2}-m}},$$ where $A=3\alpha$ is a positive constant
which quantifies if viscous effects appeared, $B$ is an arbitrary integration
constant. For small value of the scale factor $a(t)$ we have $$\rho(a) \simeq
\frac{B^{{1}/{(\frac{1}{2}-m)}}}{a^{3(1+\gamma)}},$$ which corresponds to the
universe dominated by matter satisfying equation of state $p = \gamma \rho$.
Also for a large value of scale factor (if $1+\gamma>0$) $$\rho \simeq
\bigg(\frac{A}{1+\gamma}\bigg)^{\frac{1}{\frac{1}{2}-m}} \qquad
\textrm{and}\qquad p \simeq
-\bigg(\frac{A}{1+\gamma}\bigg)^{\frac{1}{\frac{1}{2}-m}} = -\rho,$$ which
corresponds to an otherwise empty universe with cosmological constant
$\big(\frac{A}{1+\gamma}\big)^{{1}/{(\frac{1}{2}-m)}}$. For an accelerating
universe, the deceleration parameter must be negative, therefore $$
a^{3(1+\gamma)(\frac{1}{2}-m)} > \frac{B(1+\gamma)(1+3\gamma)}{3A}. $$ The above
expression indicates that if $(\frac{1}{2}-m)(1+\gamma)>0$ then a universe
starts to accelerate when the scale factor is reaching its critical value
$$a_{\text{trans}}=\bigg[\frac{B(1+\gamma)(1+3\gamma)}{2A}\bigg]^{\frac{1}{3({1}/{2}-m)(1+\gamma)}}.$$
Hence the interpolation of early matter domination phase and cosmological
constant epoch is achieved for viscous fluid when the viscosity decreases when
density is decreasing. If we put $a=1$ at the present epoch then present
Universe is accelerating if $\big(\frac{B}{A}\big)(1+\gamma)(1+3\gamma)<1$.

An important test allows to verify if the viscosity fluid may represent dark 
energy is the comparison with the supernovae type Ia data. For this purpose we 
have to calculate the luminosity distance in the model
\begin{equation}
\label{luminosity_dist}
d_L(z) =  (1+z) \frac{c}{H_0} \int_0^z \frac{\ud z'}{H(z')}.
\end{equation}

The Friedman equation can be rearranged to the form giving explicitly the
Hubble function $H(z)= {\dot a}/a$
\begin{equation}
\label{Hubble}
H(z)^2 =  H_0^2 \left[ \Omega_{\text{m}} (1+z)^3 + \Omega_{\text{visc}}
\left(A_s + (1 - A_s)(1+z)^{3(1+ \alpha)} \right)^{\frac{1}{1+\alpha}} \right]
\end{equation}
where the quantities $\Omega_i$, $i= \text{m},\text{visc},k$ represent 
fractions of critical density currently contained in energy densities of 
respective components and $\Omega_{\text{m}} + \Omega_{\text{visc}} + \Omega_k 
=  1$.

This model adds to our previous discussion of a viscous fluid
the additional elements of curvature and matter scaling like dust. As discussed
above (see equation (\ref{eq:13})), the viscous
fluid has a Chaplygin-like dependence on scale factor only in the flat dust-free
case; in other circumstances, 
equation (\ref{Hubble}) does not represent the solution for a universe containing in part
a viscous fluid. Therefore, this model is in general to be regarded as an
empirical construct. Such a model can nevertheless be compared to
data, and we do this below; where we mention model testing, we actually mean the
estimation of the $m$ and $A_{s}$ parameters
for the best fit FRW cosmological model. We will mostly use the flat model 
$k = 0$
(the exception takes place when we relax a flat prior) since the evidence of 
this case is very strong in the light of current CMBR data. 

Formula (\ref{LD}) is the most general one in the framework of the 
Friedmann-Robertson-Walker models with this unified macroscopic (phenomenological) 
description of both dark energy and dark matter (quartessence). We assume that 
$\Omega_{\text{m,0}} = \Omega_{\text{b},0}$

Finally the luminosity distance reads
\begin{equation}
d_L(z) =  (1+z) \frac{c}{H_0} \int_0^z \frac{\ud z'}{\sqrt{\Omega_{\text{m},0} 
(1+z)^3 + \Omega_{\text{visc},0}\left(A_s + (1 - A_s)(1+z)^{3(1+ \alpha)} 
\right) ^\frac{1} {1+\alpha} }}.
\label{LD}
\end{equation}

To proceed with fitting the SNIa data we need the magnitude-redshift relation
\begin{equation}
\label{m-z}
m(z,{\cal M}, \Omega_{\text{m},0}, \Omega_{\text{visc},0};A_s, \alpha) =
{\cal M} + 5 \log_{10} D_L (z, \Omega_{\text{m},0}, \Omega_{\text{visc},0};A_s, 
\alpha)
\end{equation}
where:
$$ D_L (z, \Omega_{\text{m},0}, \Omega_{\text{visc},0};A_s, \alpha) =
H_0 d_L (z, H_0, \Omega_{\text{m},0}, \Omega_{\text{Ch},0};A_s, \alpha) $$
is the luminosity distance with $H_0$ factored out so that marginalization over
the intercept
\begin{equation}
\label{intercept}
{\cal M} =  M - 5 \log_{10} H_0 +25
\end{equation}
leads actually to joint marginalization over $H_0$ and $M$ ($M$ being the
absolute magnitude of SNIa).

\begin{figure}
\includegraphics[scale=1]{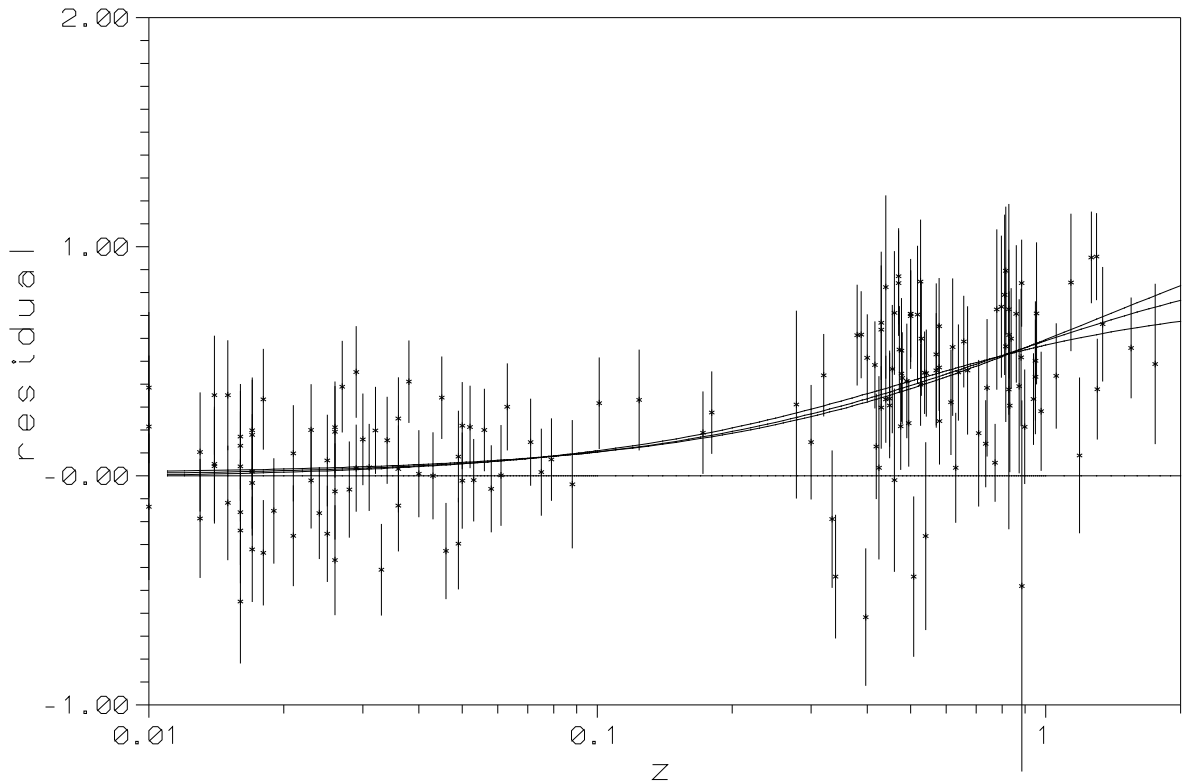}
\caption{Results from the Gold sample. Residuals (in mag) between the
Einstein-de Sitter model (zero line), and three flat models: with $m=0$, the flat
$\Lambda$CDM model ($m=-0.5$), and $m=-1.5$ (equivalent to Chaplygin gas model
with $\alpha=1$). Note that in the viscous cosmology distant $z>1$ supernovae
should be brighter than in the $\Lambda$CDM model. Therefore future distant
supernovae data can definitely distinguish: which cosmology, with dark energy
with dissipation or not?}
\label{fig:11}
\end{figure}

Then we can obtain the best fitted model minimizing the $\chi^2$ function $$
\chi^2 =  \sum_i \frac{(m_i^{Ch} - m_i^{obs})^2}{\sigma_i^2} $$ where the sum
is over the SNIa sample and $\sigma_i$ denote the (full) statistical error of
magnitude determination. This is illustrated in Fig.~\ref{fig:11} of residuals
(with respect to the Einstein-de Sitter model) and $\chi^2$ levels in the $(A_s, m)$
plane. One of the advantages of residual plots is that the intercept of the $m
- z$ curve gets cancelled. The assumption that the intercept is the same for
different cosmological models is legitimate since ${\cal M}$ is actually
determined from the low-redshift part of the Hubble diagram which should be
linear in all realistic cosmologies.

The best fit values alone are not relevant if not supplemented with the
confidence levels for the parameters. Therefore, we performed the estimation of
model parameters using the minimization procedure, based on the likelihood
function. We assumed that supernovae measurements came with uncorrelated
Gaussian errors and in this case the likelihood function ${\cal L}$ could be
determined from chi-square statistic ${\cal L} \propto \exp{(-\chi^2/2)}$
\cite{riess:1998,perlmutter:1999}.

\begin{figure}
\includegraphics[scale=0.9]{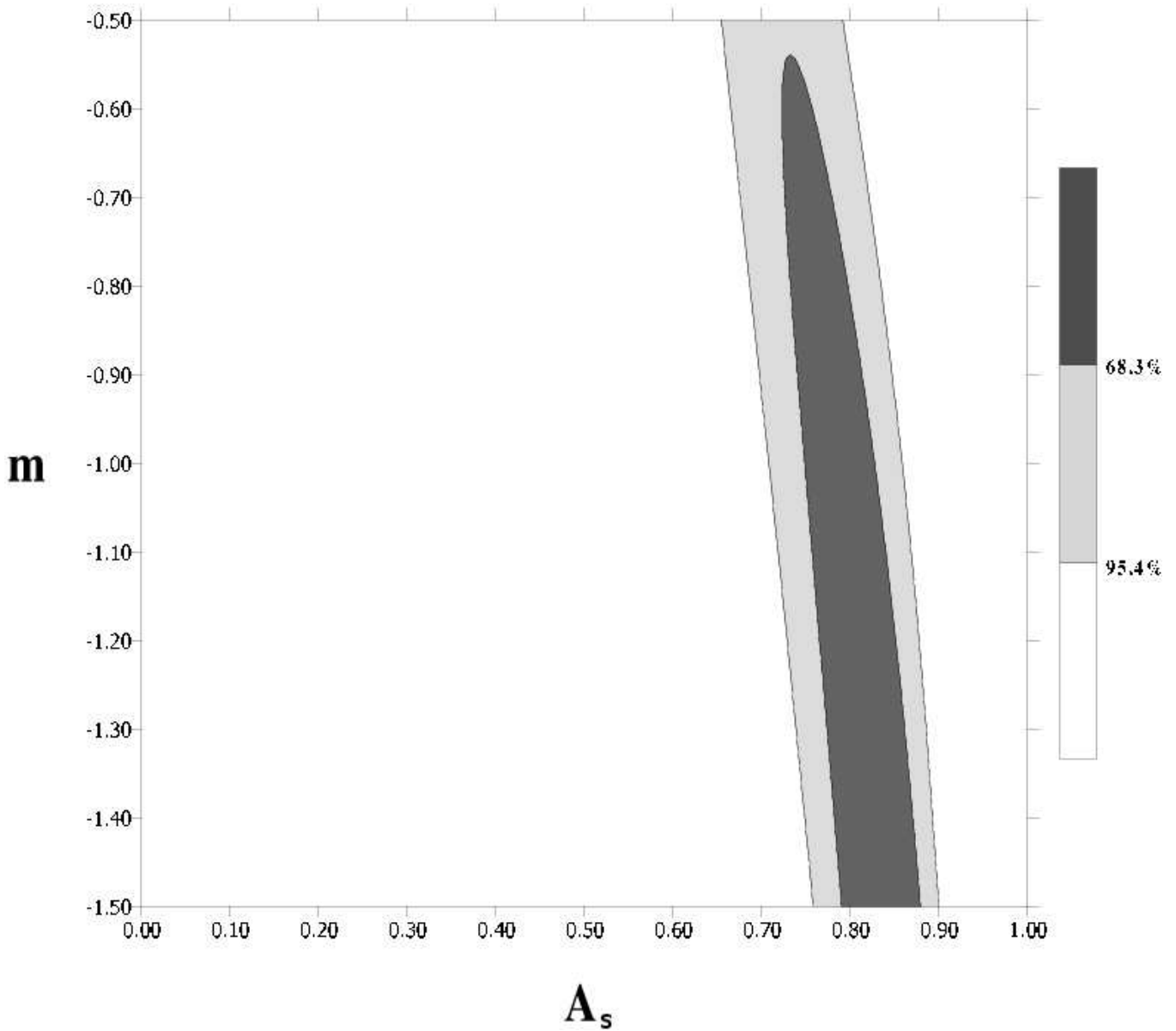}
\caption{ Confidence levels on the $(A_s,m)$ plane for the viscous
cosmological model, marginalized over ${\cal M}$, and $\Omega_{\text{m},0}$.}
\label{fig:10}
\end{figure}

Therefore we supplement our analysis with confidence intervals in the $(A_s,
m)$ plane by calculating the marginal probability density functions $$ {\cal
P}(A_s, \alpha) \propto \int \exp{\bigg(-\frac{1}{2}\chi^2(\Omega_{\text{m},0},
\Omega_{\text{visc},0}, A_s, \alpha, {\cal M})\bigg)} \ud{\cal M} $$ with 
$\Omega_{\text{m},0}, \Omega_{\text{visc},0}$ fixed ($\Omega_{\text{m},0} =  
0.0, 0.05, 0.3 $) and 
$$ {\cal P}(A_s,\alpha) \propto 
\int \exp{\bigg(-\frac{1}{2}\chi^2(\Omega_{\text{m},0}, \Omega_{\text{visc},0},
A_s, \alpha, {\cal M})\bigg)} \ud{\Omega_{\text{m},0}} $$ 
with ${\cal M}$ fixed (${\cal M} = -3.39$) respectively (a proportionality 
sign means equal up to the normalization constant) (Fig.\ref{fig:10}). In order to complete the 
picture we have also derived one-dimensional probability distribution functions 
for $\Omega_{\text{visc}}$ obtained from joint marginalization over $m$ and 
$A_s$. The maximum value of such PDF informs us about the most probable value 
of $\Omega_{\text{visc},0}$ (supported by supernovae data) within the full 
class of viscous cosmological models.

Adopting an analogy to the models with generalized Chaplygin gas (GCG) one can 
see that $\rho_{\text{visc},0}=\big(\frac{A}{1+\gamma}+B\big)^{\frac{1}{1/2-m}}$
represents the current energy density of the viscous fluid for which one can
define density parameters in the standard way. One can also calculate the
adiabatic speed of sound squared for viscous fluid
\[
c_{s}^{2}=\frac{(-\frac{1}{2}-m)\frac{A}{1+\gamma}}{\frac{A}{1+\gamma}+
\frac{B}{a^{3(1/2-m)}}}.
\]
Therefore $c_{s,0}^{2}=\frac{\alpha\frac{A}{1+\gamma}}{\frac{A}{1+\gamma}+B}$
(velocity of light at present epoch) and constants $A$ and $B$ can be expressed
in terms of quantities having well defined physical meaning. Instead of
constants $A$ and $B$ it is useful to consider new ones
$A_{s}=\frac{A}{1+\gamma}$, $B=1-A_{s}$ and $\rho_{\text{visc},0}$. Then
$c_{s,0}^{2}=(-\frac{1}{2}-m)A_{s}$ in the units of the speed of light $c$.

\subsection{Fits to $A_s$ and $m$ parameters}

\subsubsection{Samples used}

We use data sets which were compiled by Riess et al. \cite{Riess:2004}, they have 
been used by many researchers as a standard dataset. They improved the former Riess
et al. sample and discovered 16 new type Ia supernovae. It should be noted that 6 of
these objects have $z>1.25$ (out of total number of 7 object with so high red
shifts). Moreover, they compiled a set of previously observed SNIa relying on
large, published samples, whenever possible, to reduce systematic errors from
differences in calibrations. Thanks to this enriched sample it became possible to
test our prediction that distant supernovae should be brighter in viscous
cosmology than in the $\Lambda$CDM model (see discussion below).

The full Riess sample contains 186 SNIa (``Silver'' sample). Taking into
account the quality of the spectroscopic and photometric record for individual 
supernovae, they also selected a more restricted ``Gold'' sample of 157 
supernovae. We have separately analyzed the $\Lambda$CDM model for supernovae 
with $z<1$ and for all SNIa belonging to the Gold sample.

\subsubsection{Viscous cosmological models tested}

Using these samples we have tested viscous cosmology in three different classes 
of models with (1) $\Omega_{\text{m},0}= 0.3$, $\Omega_{\text{visc},0}= 0.7$; 
(2) $\Omega_{\text{m},0}= 0.05$, $\Omega_{\text{visc},0}= 0.95$ and (3) 
$\Omega_{\text{m},0} = 0$, $\Omega_{\text{visc},0} =  1$. We started with a 
fixed value of ${\cal M} = -3.39$ modifying this assumption accordingly
while analyzing different samples.

The first class was chosen as representative of the standard knowledge of
$\Omega_{\text{m},0}$ (baryonic plus dark matter in galactic halos 
\cite{peebles:2003}) with viscosity responsible for the missing part of closure 
density (the dark energy).

In the second class we have incorporated (at the level of $\Omega_{\text{m},0}$) 
the prior knowledge about the baryonic content of the Universe (as inferred 
from the BBN considerations). Hence this class is representative of the models 
in which the viscous fluid is allowed to clump and is responsible both for 
dark matter in halos as well as its diffuse part (dark energy).

The third class is a kind of toy model -- the FRW universe filled completely
with viscous fluid. We have considered it mainly in order to see how sensitive
the SNIa test is with respect to parameters identifying the cosmological
model.

Finally, we analyzed the data without any prior assumption about 
$\Omega_{\text{m},0}$.

\subsubsection{Results}

For statistical analysis we restricted the values of the $A_s$ parameter
$[0,1.0]$ and $m$ to the interval $[-0.5,-1.5]$ because of the relation
$c_s^2=\alpha \,A_s$ (where $\alpha=-0.5-m$). This constraint guarantee that in
all cases $c_s$ is real and does not exceed $c=1$. We separately analyze the case of
$m=0$, which corresponds to constant viscosity coefficient.

The results (best fits) of two fitting procedures performed on Riess samples
and with different prior assumptions concerning the cosmological models are
presented in Tables~\ref{tab:2} and \ref{tab:3}. Table~\ref{tab:2} refers to
the $\chi^2$ method whereas in Table~\ref{tab:3} the results from marginalized
probability density functions is displayed. In both cases we obtained different
values of ${\cal M}$ for each analyzed sample. In Table~\ref{tab:4} we gather
results of the statistical analysis of the viscous cosmological model performed
on the Gold Riess sample of SNIa as a minimum $\chi^{2}$ best fit for different
values of fixed $m$.

First, we present residual plots of redshift-magnitude relations between the
Einstein-de Sitter model (represented by zero line) and models with $m=0$ ---
upper curve, $m=-0.5$ (this model is equivalent to the $\Lambda$CDM model) --- 
middle curve, and $m=-1.5$ (this model is equivalent to the GCG) --- lower curve. 
One can observe that systematic deviation between these models is larger at 
higher red-shifts. The viscous model ($m=-1.5$) predicts that high redshift 
supernovae should be brighter than what is predicted with the $\Lambda$CDM 
model, while in the model with $m=0$ high redshift supernovae should be fainter 
than those predicted with the $\Lambda$CDM model.

The Riess sample leads to the results which are similar to these obtained
recently with the Astier sample \cite{Astier:2005qq}. For the Gold sample, the 
joint marginalization over parameters gives the following results: 
$\Omega_{\text{visc},0} = 1.00$ (hence $\Omega_{\text{m},0} =  0.0$), with the 
limit $\Omega_{\text{visc},0} \geq 0.80$ at the confidence level of $68.3 \%$ 
and $\Omega_{\text{visc},0} \geq 0.69$ at the confidence level of $95.4 \%$. 
$(m = -1.5, A_s =  0.83)$ with the interval $m \in (-0.86,-1.5)$ and 
$A_s \in (0.76, 0.94)$ at the confidence level of $68.3 \%$ and
$m \in (-0.55,-1.5)$ and $A_s \in (0.72,1.00)$ at the confidence level of $95.4
\%$.

\subsection{Information criteria}

In the modern observational cosmology the so called degeneracy
problem is present: many models with dramatically different scenarios are in good
agreement with the present observations. Information criteria of the model
selection \cite{liddle:2004} can be used to solve this degeneracy among some
subclass of dark energy models. Among these criteria the Akaike information
criterion (AIC) \cite{akaike:1974} and the Bayesian information criterion (BIC)
\cite{schwarz:1978} are most popular. From these criteria we can determine the
number of the essential model parameters providing the preferred fit to the
data.

The AIC is defined in the following way \cite{akaike:1974}
\begin{equation}
\label{eq:25}
\mathrm{AIC} = - 2\ln{\mathcal{L}} + 2d
\end{equation}
where $\mathcal{L}$ is the maximum likelihood and $d$ is a number of the model
parameters. The best model with a parameter set providing the preferred fit to
the data is the one that minimizes the AIC.

The BIC introduced by Schwarz \cite{schwarz:1978} is defined as
\begin{equation}
\label{eq:26}
\mathrm{BIC} = - 2\ln{\mathcal{L}} + d\ln{N}
\end{equation}
where $N$ is the number of data points used in the fit. While the AIC tends to
favor models with large number of parameters, the BIC penalizes them more 
strongly, so the BIC provides a useful approximation to full evidence in
the case of no prior on the set of model parameters \cite{parkinson:2005}.

The effectiveness of using these criteria in the current cosmological
applications has been recently demonstrated by Liddle \cite{liddle:2004} who,
taking CMB WMAP data \cite{bennett:2003}, found the number of essential
cosmological parameters to be five. Moreover he obtained the important
conclusion that spatially-flat models are statistically preferred to close
models as it was indicated by the CMB WMAP analysis (their best-fit value is
$\Omega_{tot,0} \equiv \Sigma_i \Omega_{i,0} = 1.02 \pm 0.02$ at $1\sigma$
level).

In the paper of Parkinson et~al. \cite{parkinson:2005} the usefulness of
Bayesian model selection criteria in the context of testing for double
inflation with WMAP was demonstrated. These criteria was also used recently by
us to show that models with the big-bang scenario are rather preferred over the
bouncing scenario \cite{szydlowski:2005}.

Please note that both information criteria values have no absolute sense and
only the relative values for different models are statistically interesting.
For the BIC a difference of $2$ is treated as a positive evidence ($6$ as a
strong evidence) against the model with larger value of the BIC
\cite{jeffreys:1998,mukherjee:1998}. Therefore one can order all models which
belong to the ensemble of dark energy models following the AIC and BIC values.
If we do not find any positive evidence from information criteria the models
are treated as identical and eventually additional parameters are treated as
not statistically significant. Therefore the information criteria offer the
possibility of introducing relation of weak order in the considered class of
analyzed models.

\begin{table}

\caption{Results of the statistical analysis of the viscous cosmological model (with
marginalization over ${\cal M}$) performed on the Riess samples of SNIa (Silver,
Gold) as a minimum $\chi^2$ best fit. First rows for each sample refer to no
prior on $\Omega_{\text{m},0}$. The same analysis was repeated with fixed 
priors $\Omega_{\text{m},0}=0.0$, $\Omega_{\text{m},0}=0.05$ and 
$\Omega_{\text{m},0}=0.3$.}
\begin{tabular}{@{}p{1.5cm}rrrrrrr}
\hline \hline
sample & $\Omega_{\text{m},0}$ & $\Omega_{\text{visc},0}$ & $A_s$ & $m$ & $\mathcal{M}$ & $\chi^2$ \\
\hline
Silver&  0.00 & 1.00 & 0.82 &-1.50 &15.945&229.4     \\
      &  0.00 & 1.00 & 0.82 &-1.50 &15.945&229.4     \\
      &  0.05 & 0.95 & 0.85 &-1.50 &15.945&229.6     \\
      &  0.30 & 0.70 & 0.99 &-1.50 &15.965&232.3     \\
\hline
Gold   &  0.00 & 1.00 & 0.81 &-1.50 &15.945&173.7     \\
       &  0.00 & 1.00 & 0.81 &-1.50 &15.945&173.7     \\
       &  0.05 & 0.95 & 0.84 &-1.50 &15.945&173.8     \\
       &  0.30 & 0.70 & 0.99 &-1.50 &15.965&175.6     \\
\hline
\end{tabular}
\label{tab:2}
\end{table}
				       
\begin{table}
\caption{viscous cosmological model parameter values obtained from the marginal
probability density functions calculated on the Riess samples. First rows for each
sample refer to no prior on $\Omega_{\text{m},0}$. The same analysis was 
repeated with fixed priors $\Omega_{\text{m},0}=0.0$, $\Omega_{\text{m},0}=0.05$ 
and $\Omega_{\text{m},0}=0.3$.}
\label{tab:3}
\begin{tabular}{@{}p{1.5cm}rrrrr}
\hline
\hline sample & $\Omega_{\text{m},0}$ & $\Omega_{\text{visc},0}$ &  $A_s$ & $m$ & $\mathcal{M}$\\
\hline
Silver&$0.00^{+0.18}$ & $1.00_{-0.18}$   & $0.84^{+0.09}_{-0.06}$ & $-1.5^{0.59}$ & $15.945^{+0.02}_{-0.02}$\\
      &$ 0.00$ & $1.00$ & $0.79^{+0.03}_{-0.05}$ & $-1.5^{0.52}$ & $15.955^{+0.02}_{-0.03}$\\
      &$ 0.05$ & $0.95$ & $0.81^{+0.04}_{-0.04}$ & $-1.5^{0.54}$ & $15.955^{+0.02}_{-0.02}$\\
      &$ 0.30$ & $0.70$ & $0.99^{+0.01}_{-0.03}$ & $-0.5_{-0.64}$ & $15.965^{+0.03}_{-0.02}$\\
\hline
Gold&  $0.00^{+0.20}$ & $1.00_{-0.20}$   & $0.83^{+0.11}_{-0.07}$ & $-1.5^{+0.64}$ & $15.955^{+0.03}_{-0.03}$\\
       &$ 0.00$ & $1.00$ & $0.77^{+0.04}_{-0.05}$ & $-1.5_{0.58}$ & $15.955^{+0.02}_{-0.03}$\\
       &$ 0.05$ & $0.95$ & $0.80^{+0.04}_{-0.05}$ & $-1.5_{0.59}$ & $15.955^{+0.02}_{-0.03}$\\
       &$ 0.30$ & $0.70$ & $0.99^{+0.01}_{-0.04}$ & $-0.5^{-0.64}$ & $15.965^{+0.02}_{-0.02}$\\
\hline
\end{tabular}
\end{table}

\begin{table}
\caption{Results of the statistical analysis of the viscous cosmological model 
(with marginalization over ${\cal M}$) performed on the Gold Riess sample of
SNIa as a minimum $\chi^2$ best fit for different values of fixed $m$.}
\label{tab:4}
\begin{tabular}{@{}p{1.5cm}rrrrrr}
\hline
\hline sample & $\Omega_{\text{m},0}$ &$\Omega_{\text{visc},0}$ & $A_s$ & $m$ & $\mathcal{M}$ & $\chi^2$ \\
\hline
Gold   &  0.05 & 0.95 & 0.65 & 0.0 &15.965&177.9     \\
       &  0.05 & 0.95 & 0.73 &-0.5 &15.955&175.9     \\
       &  0.05 & 0.95 & 0.79 &-1.0 &15.955&174.6     \\
       &  0.05 & 0.95 & 0.84 &-1.5 &15.945&173.8     \\
\hline
\end{tabular}
\end{table}

\begin{table}
\caption{Results of the AIC and BIC for models with different values of $m$ (with
marginalization over ${\cal M}$) performed on Gold Riess sample of SNIa. Note
that if the $m$ parameter is fixed by physics of the viscous process then viscous
model is preferred over the $\Lambda$CDM one.}
\label{tab:5}
\begin{tabular}{@{}p{1.5cm}rrr}
\hline
\hline sample & $m$ &AIC & BIC \\
\hline
Gold    & 0.0 & 183.9 & 188.0 \\
        &-0.5 & 179.9 & 186.0 \\
        &-1.0 & 178.6 & 184.7 \\
        &-1.5 & 177.9 & 183.9 \\
\hline
\end{tabular}
\end{table}
					     
In Table \ref{tab:5} we present results of analysis of four flat dark energy 
models with two free parameters, i.e., the
models with $m=0, -0.5, -1 -1.5$. On could note that both the AIC and BIC prefer
the case $m=-1.5$ i.e viscosity model with $m=-1.5$ over the $\Lambda$CDM model. It is
interesting to observe that if the value of $m$ parameter was derived from 
physics then number of independent model parameter would be lower by one and 
the viscous model is favored over the $\Lambda$CDM model by model selection
criteria (the values of AIC and BIC for the flat $\Lambda$CDM model are $179.9$ and
$186.0$, respectively, see \cite{Szydlowski:2005xv}).

\subsubsection{Bayes factor}

For deeper statistical analysis it is used the Bayes factor \cite{Trotta:2005ar,
Szydlowski:2006pz}. Using this technique we compare the viscous models with the 
concordance $\Lambda$CDM model. 

In the Bayesian framework to compare models (the model set $\{M_{i}\}$, 
$i=1,\dots,K$) is to find the value of probability in the light 
of data (so called a posterior probability) for each model. We can define the 
posterior odds for models $M_{i}$ and $M_{j}$, which (in the case when no model 
is favored a priori) is reduced to the marginal likelihood ($E$) ratio (so 
called the Bayes factor -- $B_{ij}$)
\begin{equation}\label{eq:27}
B_{ij}=\frac{\int L(\bar{\theta}|D,M_{i})P(\bar{\theta}|M_{i}) 
d \bar{\theta} }{\int L(\bar{\eta}|D,M_{j})P(\bar{\eta}|M_{j}) 
d \bar{\eta}}=\frac{E_{i}}{E_{j}},
\end{equation}
where $\bar{\theta}$ is the parameter vector, which defines model $i$, 
$L(\bar{\theta}|D,M_{i})$ is the likelihood under a model $i$, 
$P(\bar{\theta}|M_{i})$ is the prior probability for ${\bar{\theta}}$ under 
a model $i$.

It is interesting to compare the flat FRW model with bulk viscosity
and fixed baryonic matter ($\Omega_{\text{m},0} = \Omega_{\text{b},0} = 0.05$) 
with the concordance $\Lambda$CDM model. 

To compare models $M_{i}$ and $M_{j}$ one can compute $2 \ln B_{ij} = 
- (\text{BIC}_{i} - \text{BIC}_{j}) \equiv - \Delta \text{BIC} _{ij}$ which can 
be interpret as `strength of evidence' against $j$ model: $0\leq 2\ln B_{ij} < 
2$--not worth more than a bare mention, $2\leq 2\ln B_{ij} < 6$ -- positive, 
$6\leq 2\ln B_{ij} < 10$ -- strong, and $2\ln B_{ij}\ge 10$ -- very strong.

It is useful to choose one model from our model set (a reference model--$s$) and 
compare the rest models with this one model, situation in which $2\ln B_{si}>0$ 
indicates evidence against model $i$ with respect to the reference model, 
whereas $2\ln B_{si}<0$ denotes evidence in favor of model $i$.

We can compute posterior probability for model $i$ in the following way 
\begin{equation}\label{eq:28}
P(M_{i}|D)=\frac{B_{is}}{\sum _{k=1}^{K} B_{ks}}, 
\end{equation}
where $B_{is}=\exp[\frac{1}{2}\Delta BIC_{si}]$. Then one can see how prior 
believe about model probability $P(M_{i})=\frac{1}{K}$ change after inclusion 
data to analysis. This is the probability for model $i$ being the best model 
from the set of models under consideration. 

Assuming equal priors for both concurrence models $0.5$ we can calculate 
the posterior probabilities as $0.7408$ for the viscous model and $0.2592$ 
for the $\Lambda$CDM model. This indicates that the former model should be 
treated seriously as a candidate for dark energy description (see also 
\cite{Szydlowski:2006ay}).

\section{Conclusions}
In this paper we apply the theory of qualitative investigations of differential
equations to the study of dissipative cosmological model with dark energy. We
show that application of qualitative theory of dynamical system allows to
reveal some structural stability properties of this model.

We have developed a phenomenological unified model for dark energy and dark
matter through the dissipation effects acting in the flat FRW model.

We demonstrate that they describe a smooth transition from a decelerated
expansion phase dominated by matter contribution to the present dark energy
epoch. We can find one-to-one correspondence between models with generalized
Chaplygin gas ($p= -A/\rho^{\alpha}$) and models filled by viscous fluid ($p =
\gamma \rho - 3 \xi(\rho) H$) in the flat case. It is established if we use
Belinskii power law parameterization of viscosity coefficient $\xi(\rho)
\propto \rho^{m}$ and $1+\alpha=\frac{1}{2}-m$.

We conclude that while the $\Lambda$CDM model comes in
good agreement with other observational data, mainly those from WMAP and
dynamics of clusters of galaxies, the fitting quality for dissipative model is
comparable to that value obtained for models with $\Lambda$.

We supplemented our analysis with confidence intervals in the $(A_{s},m)$ plane.
Our result show also that viscous cosmology predicts that at distance $z>1$
supernovae should be brighter than in the $\Lambda$CDM model.

For deeper analysis of statistical results and to decide which model (with dark
energy with or without dissipation) is distinguished we use the Akaike and 
Bayesian information criteria. Applying the model selection criteria we show 
that both the AIC and BIC indicate that additional contribution arising from 
the dissipative effects should be incorporated to the model if only physics fix 
the value of the parameter $m$.

Since classical papers of Murphy \cite{Murphy:1973} and Heller et al. 
\cite{Heller:1973} it is well known that bulk viscosity can produce
cosmological models without the initial singularity for flat universes. In an 
analogous way one can consider influence of bulk viscosity effects on avoiding
the singularities in the future, namely so called big-rip singularities
appearing in phantom cosmology \cite{Brevik:2005}. Let us note that if 
$A_{s}<0$ ($\gamma<-1$ and $A>0$) then we have upper bound on the value of 
the scale factor if only $m>1/2$. This means that viscosity effects acting in 
phantom's matter can give rise to avoidance of big-rip singularity in the 
future.

Using the analogy of dissipative cosmology to the conservative one with the 
Chaplygin gas, one can find at least three significant features of dissipative 
cosmology \cite{Gorini:2004by}. First, similarly to the conservative models with 
the Chaplygin gas they describe a smooth transition from a decelerated to the 
present accelerated expansion. Second, the models attempt to give a unified 
phenomenological description of both dark energy and dark matter in a natural 
way without references to effects of immersion of our universe into higher 
dimensional bulk spaces or tachyon cosmological models. Third, they are a 
simple and natural extension of the $\Lambda$CDM model which has the property 
of flexibility with regard to the observational data.

The main conclusions from the dynamical systems analysis are the following
\begin{itemize}
\item The general FRW model with bulk viscosity parameterized by the Belinskii power-law 
parameterization is the two-dimensional dynamical system and framework of 
dynamical systems methods can be applied; we showed that the phase portrait of 
the model with bulk viscosity with parameter values obtained from statistical 
analysis is equivalent to the phase portrait of the $\Lambda$CDM model. 
\item For the special case of the flat model the effect of bulk viscosity are 
formally equivalent to the effects of generalized Chaplygin gas. We have found 
simple relation between parameter $\alpha$ from the equation of state of the 
generalized Chaplygin gas ($p=-A \rho^{-\alpha}$) and the parameter $m$ taken 
from the Belinskii parameterization of viscosity $\xi \propto \rho^{m}$.
These models are formally equivalent and cannot be distinguished using 
the kinematic cosmological tests based on the observables obtained from 
null geodesics.
\item Moreover in the case of flat models dynamics can be reduced to the 
form of conservative system of the Newtonian type.
\item In the special case of the flat model with constant viscosity we have 
formulated the theorem 1 which states that the solution for viscous model can 
be obtain from the the solution of the model without the viscosity. 
\end{itemize}

An additional argument for incorporating of the new dissipative parameter is
brought by the theory which will always favor the structurally stable models 
over fragile ones. We show that viscosity effects give rise to structurally 
stable evolutional scenario with squeezing bounce phase predicted by loop 
quantum gravity. However we must remember that flat cosmological model with 
viscosity effects is equivalent to the FRW model with the generalized Chaplygin 
gas and is not better than the $\Lambda$CDM model.

There is a class of dynamical systems of cosmological origin like the 
cosmological models of loop quantum gravity in which instead of an initial 
singularity we have a bounce \cite{Singh:2006im}. The corresponding dynamical 
systems is structurally unstable (see Section II) because of the presence of 
the center in the phase portrait. Therefore following the Peixoto theorem they 
are exceptional among the cosmological models on the plane. In the bouncing 
cosmology the acceleration of the Universe is only transient phenomenon and 
this is a non-generic case. Due to effect of bulk viscosity they become 
structurally stable models.

One can derive some philosophical conclusions from our analysis. The arrow of
time \cite{Zeh:2001} is a physical mystery because while fundamental laws of
physics are CPT invariant there is thermodynamical arrow of time prescribed by
the second law of thermodynamics and cosmological arrow of time prescribed by
expansion of the Universe. If our Universe is not flat then dissipative effects
of bulk viscosity can determine the cosmological arrow of time.

\acknowledgments{
We are very grateful to the anonymous referee for arguments and comments which
helped to clarify the text.
This work was supported by the Marie Curie Actions Transfer of Knowledge 
project COCOS (contract MTKD-CT-2004-517186).}

\end{document}